\documentclass[12pt]{iopart}

\usepackage{iopams}
\usepackage{bm}
\usepackage{graphicx,color}

\newcommand{\mod}{\mathop{\mathrm{mod}}\nolimits}
\newcommand{\sgn}{\mathop{\mathrm{sgn}}\nolimits}

\begin{document}

\title[XY model on the circle]{XY model on the circle: diagonalization,
spectrum, and forerunners of the quantum phase transition}

\author{A De Pasquale$^{1,2}$ and P Facchi $^{3,2}$}

\address{$^1$Dipartimento di Fisica, Universit\`a di Bari, I-70126 Bari, Italy}
\address{$^2$Istituto Nazionale di Fisica Nucleare, Sezione di Bari, I-70126 Bari, Italy}
\address{$^3$Dipartimento di Matematica, Universit\`a di Bari, I-70125 Bari, Italy}
\ead{antonella.depasquale@ba.infn.it}
\begin{abstract}
We exactly diagonalize the finite-size XY model with periodic boundary
conditions and analytically determine the ground state  energy.
We show that there are two types of fermions: singles and pairs,
whose dispersion relations have a completely arbitrary gauge-dependent sign.
It follows that  the ground state is determined by a competition between the vacuum
states (with a suitable gauge) of two parity sectors.
We finally exhibit some points in finite size systems  that forerun criticality.
They are associated to single Bogoliubov fermions and to the level crossings
between physical and unphysical states. In the thermodynamic limit
they approach the ground state and build up  singularities at logarithmic rates.
\end{abstract}

\section{Introduction}
The analysis of one dimensional spin chains is a useful approach to
the modeling of quantum computers \cite{nielsen}. This class of systems has been deeply
studied in the thermodynamic limit \cite{lieb,pfeuty,ModelliEsatti1}; however,
experimental and theoretical difficulties impose strong bounds on
the realization of large scale systems, and this has boosted a
high interest in finite size systems \cite{ModelliEsatti2,ModelliEsatti3,ModelliEsatti4,ModelliEsatti5}.
The investigation of the last few years has focused on entanglement
\cite{wooters,sarorev} in diverse finite-size models, by means of direct
diagonalization
\cite{tagliafinita1,tagliafinita2,tagliafinita3,tagliafinita4,canosarossignoli,plastina}. These studies were boosted by
the recent discovery that entanglement can detect the presence of
quantum phase transitions \cite{QPT1,QPT2,QPT3,QPT4,QPT5}.

In this article we exactly diagonalize the XY model with periodic
boundary conditions, describing a one dimensional chain made up of a
finite number of two level systems ($\frac{1}{2}$-spins) with
nearest neighbors coupling, in a constant and uniform magnetic
field. The XY model is a class of Hamiltonians distinguished by a
different value of the anisotropy coefficient, which introduces a
different coupling between the $x$ and the $y$ components of the
spins (in particular the isotropic case, corresponding to the case
in which the anisotropy coefficient vanishes, is known as XX model).

As for infinite chains \cite{ModelliEsatti1}, the diagonalization procedure is divided in three steps: the
Jordan-Wigner transformation, a deformed Fourier transform
(generalizing the discrete Fourier transform), and a gauge dependent
Bogoliubov transformation. After the Jordan-Wigner transformation
the Hamiltonian, expressed as a quadratic form of annihilation and
creation operators of spinless fermions, is characterized by the
presence of a boundary term \cite{lieb} whose contribution, which scales like
$O(1/N)$ in the calculation of real physical quantities, cannot be
neglected for finite size systems. However, this boundary term
vanishes in Fourier space if the discrete Fourier transform is
deformed with a local gauge coefficient, depending on the parity of
the spins anti-parallel to the magnetic field \cite{XX Model}.

There will emerge two
classes of fermions, coupled and single ones (in particular for the
XX model there are only single fermions). The last step of the
diagonalization procedure is the unitary Bogoliubov transformation,
given by a continuous rotation for fermion pairs  and by a discrete
one for single fermions. We will show that this unitary
transformation is gauge dependent, since it is given by two possible
continuous rotations for fermion pairs and by either the identity
or the charge conjugation operator for single fermions. From this it
follows that  the sign of the dispersion relation is completely
arbitrary, apart from the constraint that fermions belonging to the same
pair have the same sign.

From the arbitrariness of the Bogoliubov transformation it follows
that a possible expression for the diagonalized Hamiltonian is such
that for successive intervals of the magnetic field the vacuum
energies of the two parity sectors alternatively coincide with the
ground state and the first excited level: we will exhibit this mechanism of ``competition'' between vacua.

Finally we show that in finite size systems one can find the
``forerunners'' of the  points of quantum phase transition of the
thermodynamic systems. They are associated to single Bogoliubov fermions and
arise at the level crossings between physical and unphysical states.
At the values of the magnetic field corresponding to the forerunners
the second derivative of the ground state energy scales as
$-\log N$. Since in the XX model all Bogoliubov fermions are single,
one re-obtains the well known result that in the
thermodynamic limit the anisotropic case presents two discrete
quantum phase transitions whereas the isotropic or XX model is
characterized by a continuous one \cite{ModelliEsatti1}.

\section{The XY Hamiltonian}
We consider  $N$ spins on a circle with nearest neighbors
interaction in the $xy$ plane and with a constant and uniform
magnetic field along the $z$-axis. The Hilbert space is
$\mathcal{H}=\bigotimes_{i\in \mathbb{Z}_N}\mathfrak{h}_i$, where
$\mathfrak{h}_i\cong\mathbb{C}^2$ is the Hilbert space of a single
spin, and $\mathbb{Z}_N$, labeling the positions on the circle,  is
the ring of integers $\mod N$ with the standard modular addition and
multiplication. The XY Hamiltonian is given by
\begin{equation}\label{eq:XYhamiltonian}
H_{\gamma}(g)=-J \sum_{i \in \mathbb{Z}_{N}}\left[ g\sigma_i^z +
\left(\frac{1+\gamma}{2}\right)\sigma_i^x \sigma_{i+1}^x +
\left(\frac{1-\gamma}{2}\right)\sigma_i^y \sigma_{i+1}^y \right]\;,
\end{equation}
with
\begin{equation}
\sigma_i^{l}=1\otimes 1 \otimes \dots \otimes \sigma^{l}\otimes \dots \otimes 1,
\quad i\in\mathbb{Z}_N, \quad l\in\{x,y,z\}
\end{equation}
where $\sigma^l$ acts on the  $i$-th spin and
may be represented by the Pauli matrices,
\begin{equation}
\sigma^x=\left(\begin{array}{cc}0 & 1 \\1 & 0\end{array}\right), \quad
\sigma^y=\left(\begin{array}{cc}0 & -i \\i & 0\end{array}\right), \quad
\sigma^z=\left(\begin{array}{cc}1 & 0 \\0 & -1\end{array}\right).
\end{equation}
 $J>0$ is a constant with dimensions of energy and $g\in\mathbb{R}$ and
$\gamma\in[0,1]$ are two dimensionless parameters: the first one is
proportional to the transverse magnetic field and the second one is
the anisotropy coefficient and denotes the degree of anisotropy in
the $xy$ plane, varying from $0$ (XX or isotropic model) to $1$
(Ising model). As is well known, in the thermodynamic limit, the diagonalization of the XY
Hamiltonian is achieved by means of three transformations: the
Jordan-Wigner (JW), Fourier and Bogoliubov (BGV) transformations. We
will analyze in detail how the topology of the circle will induce a
deformation on these transformations in finite size chains.

\subsection{Jordan-Wigner and deformed Fourier transformations}
The Jordan-Wigner transformation is based on the  observation
that there exists a unitary mapping
\begin{equation}
\mathcal{U}: (\mathbb{C}^2)^{\otimes N}\to
\mathcal{F}_-(\mathbb{C}^N)
\end{equation}
between the Hilbert space of a system of $N$ spins $\mathcal{H}\cong(\mathbb{C}^2)^{\otimes N}$ and the fermion Fock space $\mathcal{F}_-(\mathbb{C}^N)$ of spinless
fermions on $N$ sites. Here,
\begin{equation}
\mathcal{F}_-(\mathfrak{h})= Q_{-}  \bigoplus_{n\geq 0} \mathfrak{h}^n,
\end{equation}
where $\mathfrak{h}^n=\mathfrak{h}^{\otimes n}$ for $n\geq 1$,
$\mathfrak{h}^0=\mathbb{C}$, and $Q_{-}$ is the projection onto the
subspace of  antisymmetric wave-functions \cite{BR}. In order to
simplify the notation, in the following we will use the above
isomorphism and will identify the two spaces $\mathcal{H}\cong
\mathcal{F}_-(\mathbb{C}^N)$ without making no longer mention to
$\mathcal{U}$. By virtue of this identification we can consider the
canonical annihilation and creation JW fermion operators \cite{JW}
\numparts
\begin{eqnarray}
c_i=\left(\prod_ {j\in\mathbb{Z}_{N}, j<i}\sigma^z_j\right)\,\sigma_i^- =\mathrm{e}^{\mathrm{i}\pi\bm{n}_{i\downarrow}}\sigma_i^- ,
\label{eq:c_j}\\
c_i^\dag=\left(\prod_ {j\in\mathbb{Z}_{N}, j<i}\sigma^z_j\right)\,\sigma_i^+
=\mathrm{e}^{\mathrm{i}\pi\bm{n}_{i\downarrow}}\sigma_i^+,\quad \forall
 i\in\mathbb{Z}_N
\label{eq:c_j dagger},
\end{eqnarray}
\endnumparts
where $\sigma_i^\pm=(\sigma_i^x\pm {\rmi} \sigma_i ^y)/2$ and
$\bm{n}_{i\downarrow}$ is the operator counting the number of
holes (or spins down) to the left of $i$
\begin{equation}
\bm{n}_{i\downarrow} =\sum_{j\in \mathbb{Z}_{N}, j<i}(1-c_j^\dag c_j).
\end{equation}

Note that the above definitions rely upon the following (arbitrary)
ordering of $\mathbb{Z}_N$: $[0]<[1]<\dots<[N-1]$, where
$[k]=k+N\mathbb{Z}$. In particular, if the choice of the successive
elements can be considered natural, and is well adapted to the
Hamiltonian (\ref{eq:XYhamiltonian}), the choice of the first
element $[0]$ is totally arbitrary and is related to the choice of a
privileged point of the circle.

The JW operators anti-commute both on site and on different sites
(see \eref{eq:JW anticomm relations}) whereas the Pauli operators
anti-commute only on the same site: $\forall i, j \in\mathbb{Z}_N$
\numparts
\begin{eqnarray}
\{c_i,c_j\}=0,\\ \{c_i^\dag,c_j^\dag\}=0,\\
\{c_i,c_j^\dag\}=\delta_{ij} \label{eq:JW anticomm relations}
\end{eqnarray}
\endnumparts
and
\numparts
\begin{eqnarray}
\{\sigma_i^\pm,\sigma_j^\pm\}=0\quad  \mathrm{for}\;  i=j,  \quad
[\sigma_i^\pm,\sigma_j^\pm]=0 \quad \mathrm{for}\;  i \neq j.
\end{eqnarray}
\endnumparts
From equations \eref{eq:c_j}-\eref{eq:c_j dagger} one sees that the
terms in the Hamiltonian describing the coupling between spins $[0]$
and $[N-1]=[-1]$, when written by means of the JW operators, is
characterized by an operator phase, at variance with  the other
coupling terms; for example the terms coupling the spins along the
$x$ axis become \numparts
\begin{eqnarray}
\sigma_j^x\sigma_{j+1}^x=c_jc_{j+1}^\dagger + c_jc_{j+1} +
c_{j+1}^\dagger c_j^\dagger + c_{j+1}c_j^\dagger\quad, \quad
\forall j \in \mathbb{Z}_N \backslash\{[-1]\},
\label{eq:sigmaxsigmax}\\
\sigma_{[-1]}^x\sigma_{[0]}^x=e^{{\mathrm{i}\pi
\left(\bm{n}_\downarrow+1\right)}}
 \left( c_{[-1]}c_{[0]}^\dagger + c_{[-1]}c_{[0]} + c_{[0]}^\dagger c_{[-1]}^\dagger + c_{[0]}c_{[-1]}^\dagger \right)
 \label{eq:sigmaXsigmaX0,_N-1},
\end{eqnarray}
\endnumparts
where the number operator,
\begin{equation}
\bm{n}_\downarrow=\sum_{j\in \mathbb{Z}_{N}}(1-c_j^\dag c_j) ,
\end{equation}
counts the total number of spins down in
the chain. This introduces some difficulties in the
diagonalization of the Hamiltonian, because its expression written in terms
of the fermion operators is characterized by the presence of a
boundary term with the same operator phase found in equation
(\ref{eq:sigmaXsigmaX0,_N-1})
\begin{eqnarray}\label{eq:fermionic XY hamiltonian}
\fl H_{\gamma}(g)=-J \Bigg\{ \sum_{j\in\mathbb{Z}_{N}}\left[ g(1-2c_jc_j^\dagger )+ c_jc_{j+1}^\dagger + c_{j+1}c_j^\dagger +\gamma(c_jc_{j+1}+c_{j+1}^\dagger c_j^\dagger )\right] \nonumber\\
 -\left({\rm e}^{{{\rm i}\pi
\bm{n}_\downarrow}}+1\right)\left[(c_{[-1]}c_{[0]}^\dagger +
c_{[0]}c_{[-1]}^\dagger )+ \gamma(c_{[-1]}c_{[0]}+ c_{[0]}^\dagger
c_{[-1]}^\dagger )\right]\Bigg\}.
\end{eqnarray}
In the thermodynamic limit
the boundary term can be neglected since it
introduces corrections of order $1/N$; the
problem is then reduced to the diagonalization of the so called
``c-cyclic'' Hamiltonian \cite{lieb} and can be easily achieved
by means of the discrete Fourier transform
\begin{equation}\label{eq:standard fourier transform}
\hat{c}_k=\frac{1}{\sqrt N}\sum_{j\in\mathbb{Z}_{N}}{\rm
e}^{-\frac{2\pi {\rm i}kj}{N}} c_j, \quad \forall
k\in\mathbb{Z}_N.
\end{equation}
Since we are interested in
finite size systems, with finite $N$, the boundary term cannot be neglected.
The main difficulty introduced by the boundary term in the
Hamiltonian (\ref{eq:fermionic XY hamiltonian}) is that it breaks
the periodicity of the JW operators, due to the arbitrary
dependence of the phase ${\rm e}^{{\rm i}\pi
\bm{n}_{i\downarrow}}$ on the ordering of the spins on the
circle. This phase clearly depends on the state the Hamiltonian
$H_\gamma$ is applied to. However, equation \eref{eq:fermionic XY
hamiltonian} can be simplified by noting that the parity of the
number of spins down,
\begin{equation}\label{eq:parity operator}
\mathcal{P}= {\rm e}^{{\rm i}\pi \bm{n}_\downarrow},
\end{equation}
is conserved
\begin{equation}\label{eq:commparity-hamiltonian}
[\mathcal{P},H_\gamma]=0,
\end{equation}
although not so the spin-down number operator
$\bm{n}_\downarrow$  itself.
Its spectral decomposition is
\begin{equation}
\mathcal{P}=\sum_{\varrho=\pm1} \varrho P_\varrho= P_+ -P_-,
\end{equation}
where
\numparts
\begin{eqnarray}
P_+=\sum_{n_\downarrow\,\rm{even}}|n_\downarrow\rangle\langle
n_\downarrow| ,\\
P_-=\sum_{n_\downarrow\,\rm{odd}}|n_\downarrow\rangle\langle
n_\downarrow|
\end{eqnarray}
\endnumparts
are the projection operators belonging to the eigenvalues
$\varrho=\pm1$ of $\mathcal{P}$ respectively, and
$|n_\downarrow\rangle$ is the eigenstate of $\bm{n}_\downarrow$ with
eigenvalue $n_\downarrow$. Since parity is conserved (equation
\eref{eq:commparity-hamiltonian}) the Hamiltonian
 can be decomposed as
\begin{equation}\label{eq:parity sectors}
H_{\gamma}=P_+H_{\gamma}P_+ + P_-H_{\gamma}P_-=H_{\gamma}^{(+)} +
H_{\gamma}^{(-)},
\end{equation}
and the analysis can be separately performed in each parity sector,
where $\mathcal{P}$ acts as a superselection charge.

In each sector the XY Hamiltonian can be diagonalized  by
deforming the discrete Fourier transform by means of a local gauge
$\alpha_j$ ($j\in \mathbb{Z}_N$),
\begin{equation}
\hat{c}_k=\frac{1}{\sqrt N} \sum_{j\in\mathbb{Z}_N}
\exp\left(-\frac{2\pi {\rm i} }{N} \left(k j + \alpha_j \right) \right) c_j,
\quad
k\in \mathbb{Z}_N.
\label{eq:deformed fourier transform dir}
\end{equation}
The inverse formula reads
\begin{equation}
c_j=\frac{1}{\sqrt N} \sum_{k\in\mathbb{Z}_N}
\exp\left(\frac{2\pi {\rm i} }{N} \left(k j + \alpha_j \right) \right) \hat{c}_k, \quad
j\in \mathbb{Z}_N.
\label{eq:deformed fourier transform}
\end{equation}
This deformation preserves the anti-commutation
relations in the Fourier space \numparts
\begin{eqnarray}\label{eq:anticomm JW relations}
\{\hat{c}_k,\hat{c}_{k'}\}=0,\\
\{\hat{c}_k^\dag,\hat{c}_{k'}^\dag\}=0,\\
\{\hat{c}_k,\hat{c}_{k'}^\dag\}=\delta_{kk'},
\end{eqnarray}
\endnumparts
$\forall k, k'\in \mathbb{Z}_N$.
The local gauge  $\exp\left(2\pi {\rm i}\alpha_j/N \right)$
can be determined by imposing that the Fourier transforms of
(\ref{eq:sigmaxsigmax}) and (\ref{eq:sigmaXsigmaX0,_N-1}) have the
same form. Considering the first terms in the sums  one gets
\numparts
\begin{eqnarray}
c_jc_{j+1}^\dag &=& {\rm e}^{\frac{2\pi {\rm i}}{N} (\alpha_j-\alpha_{j+1})}
\frac{1}{N}\sum_{k,k'\in\mathbb{Z}_{N}}
{\rm e}^{\frac{2\pi{\rm i}}{N}[jk -(j+1)k']}\hat{c}_k\hat{c}_{k'}^\dag , \\
{\rm e}^{{\rm i}\pi({n_\downarrow}+1)}c_{[-1]}c_{[0]}^\dag
&=&{\rm e}^{{\rm i}\pi({n_\downarrow}+1)} {\rm e}^{\frac{2\pi{\rm
i}}{N} ( \alpha_{[-1]}-\alpha_{[0]})}
\frac{1}{N}\sum_{k,k'\in\mathbb{Z}_{N}} {\rm e}^{\frac{2\pi{\rm
i}}{N} [(-1)k - 0 k']} \hat{c}_k\hat{c}_{k'}^\dagger,
\end{eqnarray}
\endnumparts
where ${\rm e}^{{\rm i}\pi({n_\downarrow}+1)}$ is uniquely defined
in the sector; they have the same form when, $\forall
j\in\mathbb{Z}_N$,
\begin{equation}\label{eq:phase condition}
\exp\left(\frac{2\pi{\rm i}}{N}(\alpha_j-\alpha_{j+1})\right)=
\exp\left({\rm i}\pi({n_\downarrow} + 1)\right)\exp\left(\frac{2\pi{\rm
i}}{N}(\alpha_{[-1]}-\alpha_{[0]})\right).
\end{equation}
Therefore, the left hand side,  like the
right hand side, must not depend on $j$:
\begin{equation}\label{eq:alpha condition}
\alpha_{j+1}-\alpha_j=\alpha, \qquad \forall j \in \mathbb{Z}_N,
\end{equation}
with $\alpha$ solution to the equation
\begin{equation}\label{eq:fase alpha}
\exp\left(2\pi{\rm i}\alpha\right)=\exp\left({\rm
i}\pi({n_\downarrow}+1)\right),
\end{equation}
and the phase associated to the first site $\alpha_0$ completely free.
The solutions in the two parity sectors are
\begin{equation}
\label{alfadef}
\alpha\equiv \frac{1+\varrho}{4} \; (\mod N)\equiv
 \cases{0\; (\mod N) &{\rm if}   $\varrho=-1$ \quad ($n_\downarrow$ \ {\rm odd})\\
\frac{1}{2}\; (\mod N) &{\rm if}     $\varrho=+1$ \quad
($n_\downarrow$ \ {\rm even}). }
\end{equation}
Summarizing, by substituting the (sector dependent) deformed Fourier
transform
\begin{equation}
c_j=\frac{{\rm e}^{\frac{2\pi {\rm i}}{N}\alpha_0}}{\sqrt N} \sum_{k\in\mathbb{Z}_N}
\exp\left(\frac{2\pi {\rm i}j }{N}(k + \alpha)  \right) \hat{c}_k, \quad
j\in \mathbb{Z}_N,
\end{equation}
into equation \eref{eq:fermionic XY hamiltonian} we obtain
\begin{eqnarray}
\label{eq:XY hamiltonian with deformed Fourier transform}
\fl \qquad H_{\gamma}^{(\varrho)}(g)&=&-J\sum_{k\in\mathbb{Z}_{N}} \left \{
g+2\hat{c}_k\hat{c}_k^\dagger
              \left[\cos\left(2\pi\frac{\alpha+k}{N}\right)-g\right] \right. \nonumber\\
        & & \phantom{-J\sum_{k\in\mathbb{Z}_{N}} \left[ \right. }\;
        \left. + {\rm i} \gamma \sin\left(2\pi \frac{\alpha + k}{N}\right)
             \left({\rm e}^{\frac{4\pi{\rm i} \alpha_0}{N}}\hat{c}_{\bar{k}}\hat{c}_k
             +{\rm e}^{-\frac{4\pi{\rm i}\alpha_0}{N}}\hat{c}_{\bar{k}}^\dagger\hat{c}_k^\dagger\right) \right\}P_\varrho,
\end{eqnarray}
where $\forall k\in \mathbb{Z}_N$
\begin{equation}
\bar{k}= -2\alpha-k + N \mathbb{Z}.
\end{equation}
A comment is now in order. Note that, alternatively, instead of the Fourier transform one could have deformed the JW transformation in the following way
\begin{equation}
c_j= \mathrm{e}^{\mathrm{i}\pi n_{j\downarrow}}
\mathrm{e}^{-\frac{2\pi
\mathrm{i}}{N}(j \alpha+\alpha_0)}
\sigma_j^- , \quad \forall j \in
\mathbb{Z}_N,
\end{equation}
and would have obtained the same results.

\subsection{The Bogoliubov transformation} \label{sec:the bogoliubov transform}
Observe that when $\gamma >  0$ the last term in Hamiltonian (\ref{eq:XY hamiltonian with deformed Fourier transform})
 couples fermions with momenta $k$ and
$\bar{k}$. In fact, there are two types of fermions, the single and the coupled ones (fermion pairs). Their momenta $k$ belong to the two sets
\begin{eqnarray}
\mathcal{S}_\varrho=\left\{k\in\mathbb{Z}_N  \Big|  k = \bar{k}  \right\}=
\left\{k\in\mathbb{Z}_N  \Big| 2 k = - \frac{1+\varrho}{2} + N\mathbb{Z}  \right\},\\
\mathcal{C}_\varrho= \mathbb{Z}_N \backslash \mathcal{S}_\varrho,
\end{eqnarray}
respectively. Note that the mapping $k\mapsto\bar{k}$ is an
involution of $\mathbb{Z}_N$, i.e.\ $\bar{\bar{k}}=k$. Therefore it
can be viewed as an action of the group $\mathbb{Z}_2$ on the space
$\mathbb{Z}_N$. From this perspective, $\mathcal{S}_\varrho$ and
$\mathcal{C}_\varrho$ are nothing but the sets of
 points belonging to one-element and two-element orbits of the above action, respectively. The terms in the Hamiltonian involving pairs $(k,\bar{k})$ of fermions,  in fact, depend only on the orbit.
The XY Hamiltonian can be written accordingly as
\begin{eqnarray}\label{eq:XYHamiltonian after deformed fourier}
\fl  H_{\gamma}^{(\varrho)}(g)= 2J\Bigg\{\sum_{k \in
\mathcal{S}_\varrho}\left[\cos\left (2\pi\frac{\alpha+k}{N}\right)-g\right]
\left(\hat{c}_k^\dag\hat{c}_k-\frac{1}{2}\right)
 + \frac{1}{2} \sum_{k \in \mathcal{C}_\varrho} \bm{C}_k^\dagger
h_{\mathrm{\gamma}}(k)\bm{C}_k\Bigg\} P_\varrho , \quad
\end{eqnarray}
where
\begin{eqnarray}
\bm{C}_k=\pmatrix{{\rm e}^{\frac{2\pi{\rm i}\alpha_0}{N}}\hat{c}_k \cr
\cr {\rm e}^{-\frac{2\pi{\rm
i}\alpha_0}{N}}\hat{c}_{\bar{k}}^\dagger}
\end{eqnarray}
and $h_{\mathrm{\gamma}}(k)$ is an hermitian operator on $\mathbb{C}^{2}$ given by
\begin{eqnarray}\label{eq:matrix Hgamma in C4}
h{_\gamma}(k)=\pmatrix {\cos\left(2\pi\frac{\alpha+k}{N}\right)-g& &{\rm i}
\gamma\sin\left(2\pi \frac{\alpha + k}{N}\right)\cr \cr {-\rm
i}\gamma \sin\left(2\pi \frac{\alpha + k}{N}\right)&
&-\cos\left(2\pi\frac{\alpha+k}{N}\right)+g \cr}.
\end{eqnarray}
The factor $1/2$ in front of the pair terms in (\ref{eq:XYHamiltonian after deformed fourier}) derives from the identity $\bm{C}_{\bar{k}}^\dagger
h_{\mathrm{\gamma}}(\bar{k})\bm{C}_{\bar{k}}= \bm{C}_k^\dagger
h_{\mathrm{\gamma}}(k)\bm{C}_k$, that expresses the fact that the various terms depend only on the orbit they belong to.

\begin{figure}
\begin{center}
\includegraphics[width=0.5\columnwidth]{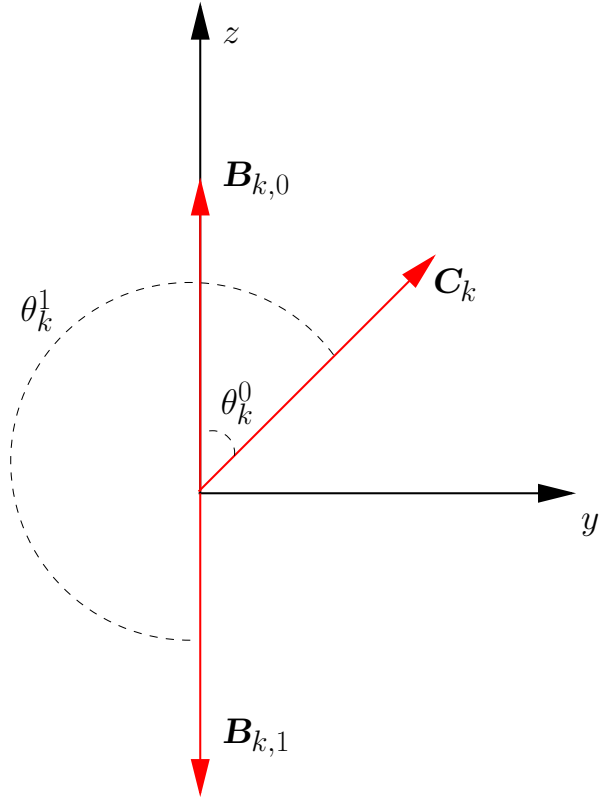}
\end{center}
\caption{Bogoliubov rotation for fermion pairs.}
\label{fig:bogoliubov}
\end{figure}

Let us  first focus on  fermion pairs. For each $k \in
\mathcal{C}_\varrho$, $h_{\gamma}(k)$ can be written as
\begin{eqnarray}
h_{\gamma}(k)= -\gamma \sin\left(2\pi \frac{\alpha + k}{N}\right)
\sigma^y + \left[\cos\left(2\pi\frac{\alpha+k}{N}\right)-g\right] \sigma^z;
\end{eqnarray}
thus $h_\gamma$ can be thought as a vector in the $yz$
plane of the internal space of the pair, and is diagonalized (i.e.\ rotated up to the $z$ direction) by a unitary rotation along $x$,
\begin{equation}
R_x(\theta_k) h_\gamma(k) R_x(\theta_k)^\dagger =  \tilde h \sigma^z,
\end{equation}
with $\tilde h\in\mathbb{R}$ and
\begin{equation}\label{eq:x axis rotation}
R_x(\theta_k)=\exp\left(-\mathrm{i}\frac{\theta_k}{2}
\sigma^x\right)=\pmatrix{\cos\frac{\theta_k}{2}&-\mathrm{i}\sin\frac{\theta_k}{2}\cr\cr-\mathrm{i}
\sin\frac{\theta_k}{2} &\cos \frac{\theta_k}{2}}.
\end{equation}
By recalling that $R_x(\theta_k)\sigma^yR_x(\theta_k)^\dagger =
\cos\theta_k\sigma^y  + \sin\theta_k \sigma^z$ and
$R_x(\theta_k)\sigma^zR_x(\theta_k)^\dagger = \cos\theta_k \sigma^z
- \sin\theta_k \sigma^y$, and by requiring that the terms
proportional to $\sigma^y$ vanish, one obtains
\begin{equation}
\gamma \sin \frac{2\pi (\alpha+k)}{N}  \cos\theta_k+ \left[\cos\frac{2\pi (\alpha+k)}{N}-g \right]\sin\theta_k=0.
\end{equation}
For each pair $(k,\bar{k})$, there are \emph{two} possible solutions that differ by $\pi$,
\begin{equation}
\theta_k^{s}=\theta_k + s \pi, \qquad \theta_{\bar{k}}^{s}=-\theta_k + s \pi,  \qquad s\in\{0,1\},
\end{equation}
where
\begin{equation}\label{eq:thetak}
\theta_k=\arctan \left(
\frac{\gamma \sin\left(2\pi\frac{\alpha+k}{N}\right)}{g-\cos\left(2\pi\frac{\alpha+k}{N}\right)}\right) \in \left(-\frac{\pi}{2},\frac{\pi}{2}\right],
\end{equation}
and
\begin{eqnarray}
\tilde h &=& - \gamma \sin \frac{2\pi (\alpha+k)}{N}  \sin\theta_k^s+ \left[\cos\frac{2\pi (\alpha+k)}{N}-g \right]\cos\theta_k^s
\nonumber\\
&=&  \left[\cos\frac{2\pi (\alpha+k)}{N}-g \right]\cos\theta_k^s \left(1+\tan^2 \theta_k^s \right).
\end{eqnarray}
The unitary transformation $R_x(\theta_k^{s})$
applied to $\bm{C}_k$ defines a new vector of fermion operators
\begin{equation}\label{eq:vettore di bogoliubov}
\bm{B}_{k,s}=
\pmatrix{b_{k} \cr \cr
b_{\bar{k}}^\dagger }=
R_x(\theta_k^{s}) \bm{C}_k, \qquad k\in\mathcal{C}_\varrho, \quad s\in\{0,1\},
\end{equation}
where $\bm{B}_{k,0}$ is related to $\bm{B}_{k,1}$ by the
relation
\begin{equation}\label{eq:trasf da theta- a theta +}
\bm{B}_{k,1}=R_x(\pi) \bm{B}_{k,0}, \qquad k\in\mathcal{C}_\varrho.
\end{equation}
See \fref{fig:bogoliubov}.
The fermion operators $b_k$ and $b_{\bar{k}}$ are
the Bogoliubov operators, and $R_x$ is the
Bogoliubov transformation for  fermion pairs. By noting that
\begin{equation}
\cos\theta_k^s = (-1)^s \left(1+\tan^2\theta_k^s \right)^{-1/2},
\end{equation}
for each pair of
momenta one gets
\begin{equation}
H_{\gamma,k}^{(\varrho)}= \bm{C}_k^\dagger
h_{\mathrm{\gamma}}(k)\bm{C}_k=(-1)^s \varepsilon_k^{(\varrho)}(g) \,
\bm{B}_{k,s}^\dag \sigma^z \bm{B}_{k,s},   \quad k\in\mathcal{C}_\varrho, \quad s\in\{0,1\},
\end{equation}
where $\varepsilon_k^{(\varrho)}$ is the dispersion relation for
fermion pairs
\begin{equation}
\fl \quad \varepsilon_k^{(\varrho)}(g)=  \sgn \! \left[\cos
\frac{2\pi(\alpha+k)}{N}-g\right]\sqrt{ \left[\cos
\frac{2\pi(\alpha+k)}{N}-g \right]^2+\gamma^2\sin^2
\frac{2\pi(\alpha+k)}{N}}.
\label{eq:dispersionrelation}
\end{equation}
Here $\sgn$ is the sign function, $\sgn x= x/|x|$ for $x\neq 0$, and $\sgn 0=0$.
We stress that for each $k\in \mathcal{C}_\varrho$ the Bogoliubov
rotation is defined independently on the other pairs, and so the sign of the dispersion relation can be chosen in a
completely arbitrary way pair by pair.
It is not difficult to show that the unitary operator on the Fock space $\mathcal{F}_-(\mathbb{C}^N)$ corresponding to a Bogoliubov rotation $R_x(\theta)$,
\begin{equation}
b_k = U_k(\theta_k^s)^\dagger \hat{c}_k U_k(\theta_k^s),  \qquad
b_{\bar{k}} = U_k(\theta_k^s)^\dagger \hat{c}_ {\bar{k}} U_k(\theta_k^s),
\end{equation}
reads
\begin{equation}\label{eq:Bog transform with second quantization}
U_k(\theta)=\exp\left(-{\rm i} \frac{\theta}{2} K_k\right),
\qquad
K_k = {\hat
{c}}_k^\dag\hat {c}_{\bar{k}}^\dag+{\hat {c}}_{\bar{k}} \hat
{c}_k, \quad k\in \mathcal{C}_\varrho.
\end{equation}
Its action on the Hamiltonian is
\begin{eqnarray}
U_k(\theta_k^s) H_{\gamma,k}^{(\varrho)}  U_k(\theta_k^s)^\dagger  =   (-1)^s \varepsilon_k^{(\varrho)} \left( \hat{c}_k^\dagger \hat{c}_k - \hat{c}_{\bar{k}} \hat{c}_{\bar{k}}^\dagger \right)  \quad k\in \mathcal{C}_\varrho.
\end{eqnarray}

Observe that since $K_k$ are quadratic with respect of creation and
annihilation operators they commute with the parity operator
(\ref{eq:parity operator}),
\begin{equation}
\mathcal{P}= U_k(\theta) \mathcal{P} U_k^\dag (\theta),  \quad k\in \mathcal{C}_\varrho,
\end{equation}
 and this means that the Bogoliubov
transformation for fermion pairs preserves the parity sector.
Finally, according to (\ref{eq:trasf da theta- a theta +}) one
gets the relation
\begin{equation}
U_k(\theta_k^1)=V_{k\bar{k}} U_k(\theta_k^0), \quad \mathrm{with}\;
V_{k\bar{k}}=U_k(\pi), \quad k\in\mathcal{C}_\varrho.
\end{equation}
Note that the unitary operator $V_{k\bar{k}}$ can be decomposed in the form
\begin{equation}
V_{k\bar{k}}=S_{k\bar{k}} C_k C_{\bar{k}}
\end{equation}
where $C_k$ and $S_{k\bar{k}}$ are respectively the charge conjugation
and the swapping operator \numparts
\begin{eqnarray}
C_k \hat{c}_k C_k^\dag=\hat{c}_k^\dag, \\
S_{k\bar{k}}\hat{c}_kS_{k\bar{k}}^\dag =- {\rm
i}\hat{c}_{\bar{k}}, \quad k\in\mathcal{C}_\varrho,
\end{eqnarray}
\endnumparts
whose explicit expressions are
\numparts
\begin{eqnarray}
C_k =\exp\left({\rm i} \frac{\pi}{2} (\hat{c}_k +
\hat{c}_k^\dag)\right),\\
S_{k\bar{k}}=\exp\left({\rm i} \frac{\pi}{2} (\hat{c}_k ^\dag
\hat{c}_{\bar{k}}  + \hat{c}_{\bar{k}}^\dag \hat{c}_k)\right),
\quad k\in\mathcal{C}_\varrho.
\end{eqnarray}
\endnumparts

Consider now the case of single fermions, $k \ \in \mathcal{S}_\varrho$.
The set $\mathcal{S}_\varrho$ depends both on the parity sector and on the parity of $N$.
For $N$ even one gets
\numparts
\begin{eqnarray}
\mathcal{S}_\varrho
=\cases{\left\{[0], \left[\frac{N}{2}\right]\right\}  &if   $\varrho=-1$,\\
\emptyset  & if $\varrho=+1$,}
\label{eq:Srho N even}
\end{eqnarray}
while, for $N$ odd,
\begin{eqnarray}
\mathcal{S}_\varrho=
\cases{\{[0]\}  & if $\varrho=-1$,\\
 \left\{\left[\frac{N-1}{2}\right]\right \}  & if $\varrho=+1$.}
\label{eq:Srho N odd}
\end{eqnarray}
\endnumparts
One can look at single fermions as a degenerate case of Bogoliubov pairs. Indeed,
\Eref{eq:thetak} reduces to
\begin{equation}
\tan \theta_k =0, \qquad k\in\mathcal{S}_\varrho,
\end{equation}
whose solutions are given by $\theta_k^{s}=s \pi$, with $s\in\{0,1\}$.
Therefore, in this case we are free to choose between two possible unitary transformation: the identity and the charge conjugation,
\begin{equation}\label{eq:bogoliubov for single fermions}
 U_k= (C_k)^0= 1, \qquad \mathrm{or}\quad  U_k= C_k, \qquad k\in\mathcal{S}_\varrho.
\end{equation}
Note that, if charge conjugation is chosen,  parity is not
preserved; rather the two parity sectors are swapped by the
Bogoliubov transformation,
\begin{equation}
\mathcal{P} = - C_k \mathcal{P} C_k^\dag .
\end{equation}
Finally, note that for single fermions the dispersion
relation (\ref{eq:dispersionrelation})  reduces to
\begin{eqnarray}
\varepsilon_k^{(\varrho)}(g)= \cos\left(2\pi\frac{\alpha+k}{N}\right)-g,
\qquad k\in\mathcal{S_\varrho},
\label{eq:dispersion relation single k}
\end{eqnarray}
since
\begin{equation}
\sin\left(2\pi\frac{\alpha+k}{N}\right)=0,\qquad k\in\mathcal{S_\varrho}.
\end{equation}

In conclusion, the total Bogoliubov transformation that diagonalizes the
Hamiltonian (\ref{eq:XYHamiltonian after deformed fourier}) has the form
\begin{equation}\label{eq:total Bogoliubov}
U_B(g,\gamma;\varrho,\bm{s})=\prod_{k \in \mathcal{C}_\varrho
/\mathbb{Z}_2} U_k(\theta_k) (V_{k\bar{k}})^{s_k} \prod_{j \in
\mathcal{S}_\varrho}(C_j)^{s_j},
\end{equation}
where
\begin{equation}
\bm{s}=(s_k) \in \{0,1\}^N, \qquad \mathrm{with}\; s_k=s_{\bar{k}},
\label{eq:constraint}
\end{equation}
and $\mathcal{C}_\varrho /\mathbb{Z}_2$ denotes that, in the case
of coupled fermions, one must consider only one element for each pair (orbit of $\mathbb{Z}_2$).
Due to the constraint in (\ref{eq:constraint}), the Bogoliubov unitary transformation
has a gauge freedom represented by the arbitrary choice of  a binary
vector of length $|\mathcal{S}_\varrho|+ |\mathcal{C}_\varrho|/2$.

Note that the anti-commutation relations are preserved by the Bogoliubov transformation,
while the parity sectors are swapped according to
\begin{equation}
\mathcal{P} = (-1)^{|\bm{s}|_\varrho} U_{B}(g,\gamma;\varrho,\bm{s})\, \mathcal{P}\,  U_{B} (g,\gamma;\varrho,\bm{s})^\dag,
\end{equation}
where
\begin{equation}
|\bm{s}|_\varrho=|\bm{s}_{\mathcal{S}_\varrho}|
=\sum_{k\in \mathcal{S}_\varrho} s_k.
\end{equation}
Therefore,  one obtains the final expression of the diagonalized Hamiltonian
\begin{eqnarray}
  \tilde{H}_{\gamma}^{(\varrho)}(g)&=&
U_{B}(g,\gamma;\varrho,\bm{s}) H_{\gamma}^{(\varrho)}(g) U_{B}(g,\gamma;\varrho,\bm{s})^\dagger
\nonumber\\
&=& 2J\sum_{k \in \mathbb{Z}_N} (-1)^{s_k}\varepsilon_k^{(\varrho)}(g) \left(\hat{c}_k^\dag \hat{c}_k -\frac{1}{2}\right)
 P_{\bar{\varrho}},
\end{eqnarray}
where
\begin{equation}
\bar{\varrho}=(-1)^{|\bm{s}|_\varrho} \varrho,
\end{equation}
which depends on an arbitrary vector $\tilde{\bm{s}}\in \{0,1\}^{|\mathcal{S}_\varrho|+ |\mathcal{C}_\varrho|/2}$, that generates $\bm{s}$ by the relation $s_k=s_{\bar{k}}=\tilde{s}_k$.
Note  that the physical part of $\tilde{H}_{\gamma}^{(\varrho)}$ acts on the sector of parity $\bar{\varrho}$.

\subsection{XY ground state: vacua competition}
\label{sec:XY vacua competition}

One can use the gauge freedom of the Bogoliubov transformation (\ref{eq:total
Bogoliubov}) in the following convenient way. Let $\bm{s}=\bm{s}(g)$ be a function of the intensity of the magnetic field $g$, such that $(-1)^{s_k(g)} \, \varepsilon_k^{(\varrho)}(g)\geq 0$ for every $g\in\mathbb{R}$.
From (\ref{eq:dispersionrelation}) this means that
\begin{equation}
(-1)^{s_k(g)} = \sgn \!\left[\cos
\frac{2\pi(\alpha+k)}{N}-g\right],
\end{equation}
that is
\begin{equation}
s_k(g) = \frac{1}{2}-\frac{1}{2}\sgn\! \left[\cos
\frac{2\pi(\alpha+k)}{N}-g\right].
\label{eq:skg}
\end{equation}
Note that since $s_{\bar{k}}(g)=s_{k}(g)$, the above solution is consistent with the constraint (\ref{eq:constraint}) of $\bm{s}$.
Therefore, the diagonalized expression of the XY
Hamiltonian reads
\begin{equation}
H_{\gamma}^{(\varrho)}(g) =2 J \sum_{k\in\mathbb{Z}_{N}}  | \varepsilon_k^{(\varrho)}(g)| \left(\hat{c}_k^\dag
\hat{c}_k -\frac{1}{2}\right) P_{\bar{\varrho}(g)}, \qquad  \bar{\varrho}(g)=(-1)^{|\bm{s}(g)|_\varrho}\varrho.
\label{eq:H vacua comp}
\end{equation}
With this choice one has that in each parity sector the lowest
energy state is the one with zero fermions (\textit{vacuum state})
whose energy density$\; /J$  is given by
\numparts
\begin{eqnarray}
E_{\rm vac}^{(-)}=-\frac{1}{N}\sum_{k\in \mathbb{Z}_{N}}\sqrt{
\left[g-\cos\left( \frac{2\pi k}{N}
\right)\right]^2+\gamma^2\sin^2\left( \frac{2\pi
k}{N}\right)}\label{eq:XYvacuum energy alpha 0},\\
E_{\rm vac}^{(+)}=-\frac{1}{N}\sum_{k\in
\mathbb{Z}_{N}} \sqrt{\left[g-\cos\left( \frac{2\pi
k}{N}+\frac{\pi}{N} \right)\right]^2+\gamma^2\sin^2\left(
\frac{2\pi k}{N}+\frac{\pi}{N}\right)}.
\label{eq:XYvacuum energy
alpha 1/2}
\end{eqnarray}
\endnumparts
Note, however, that a condition must be satisfied: the Bogoliubov
vacuum state is a physical state, \emph{provided} that it has the
right parity  $\bar{\varrho}(g)$. Were this not the case, the
projection $P_{\bar{\varrho}(g)}$ would automatically rule it out.

Let us look at the function $\bar{\varrho}(g)$ more closely. For $N$ even we have from (\ref{eq:Srho N even}) and~(\ref{eq:skg})
\begin{eqnarray}
|\bm{s}(g)|_\varrho &=& \left(s_{[0]}(g)+s_{\left[\frac{N}{2}\right]}(g)\right)\delta_{\varrho,-1}
\nonumber\\
& = & \left( 1+\frac{1}{2} \sgn(1-g) -\frac{1}{2} \sgn(1+g)\right)\delta_{\varrho,-1} ,
\end{eqnarray}
whence
\begin{eqnarray}
\bar{\varrho}(g) = (-1)^{|\bm{s}|_\varrho} \varrho = \delta_{\varrho, -1}
\sgn(1-g^2) + \delta_{\varrho, +1},
\end{eqnarray}
that is
\begin{eqnarray}
\bar{\varrho}(g) = \sgn\!\left(1-\frac{1-\varrho}{2} g^2\right)   \qquad  (N\; \mathrm{even}).
\label{eq:rhog even}
\end{eqnarray}
For $N$ odd we have from (\ref{eq:Srho N odd}) and (\ref{eq:skg})
\begin{eqnarray}
|\bm{s}(g)|_\varrho &=& s_{[0]}(g) \delta_{\varrho,-1} + s_{\left[\frac{N-1}{2}\right]}(g) \delta_{\varrho,+1}
\nonumber\\
& = & \frac{1}{2}  -\frac{1}{2} \sgn(1-g)\;\delta_{\varrho,-1} + \frac{1}{2} \sgn(1+g)\; \delta_{\varrho,+1} ,
\end{eqnarray}
whence
\begin{eqnarray}
\bar{\varrho}(g) = - \sgn(1+ \varrho g)  \qquad  (N\; \mathrm{odd}).
\label{eq:rhog odd}
\end{eqnarray}

Since the vacuum state has $N$ holes, its parity is $(-1)^N$, and it is a physical state only if
\begin{equation}
\bar{\varrho}(g)=(-1)^N .
\label{eq:physical}
\end{equation}
Equation (\ref{eq:physical}) is satisfied for arbitrary $\varrho$ when $g\in (-1,1)$, while
it is true only for $\varrho=(-1)^N$ for $g<-1$, and $\varrho=+1$ for $g>1$.
\begin{figure}
\begin{center}
\includegraphics[width=0.7\columnwidth]{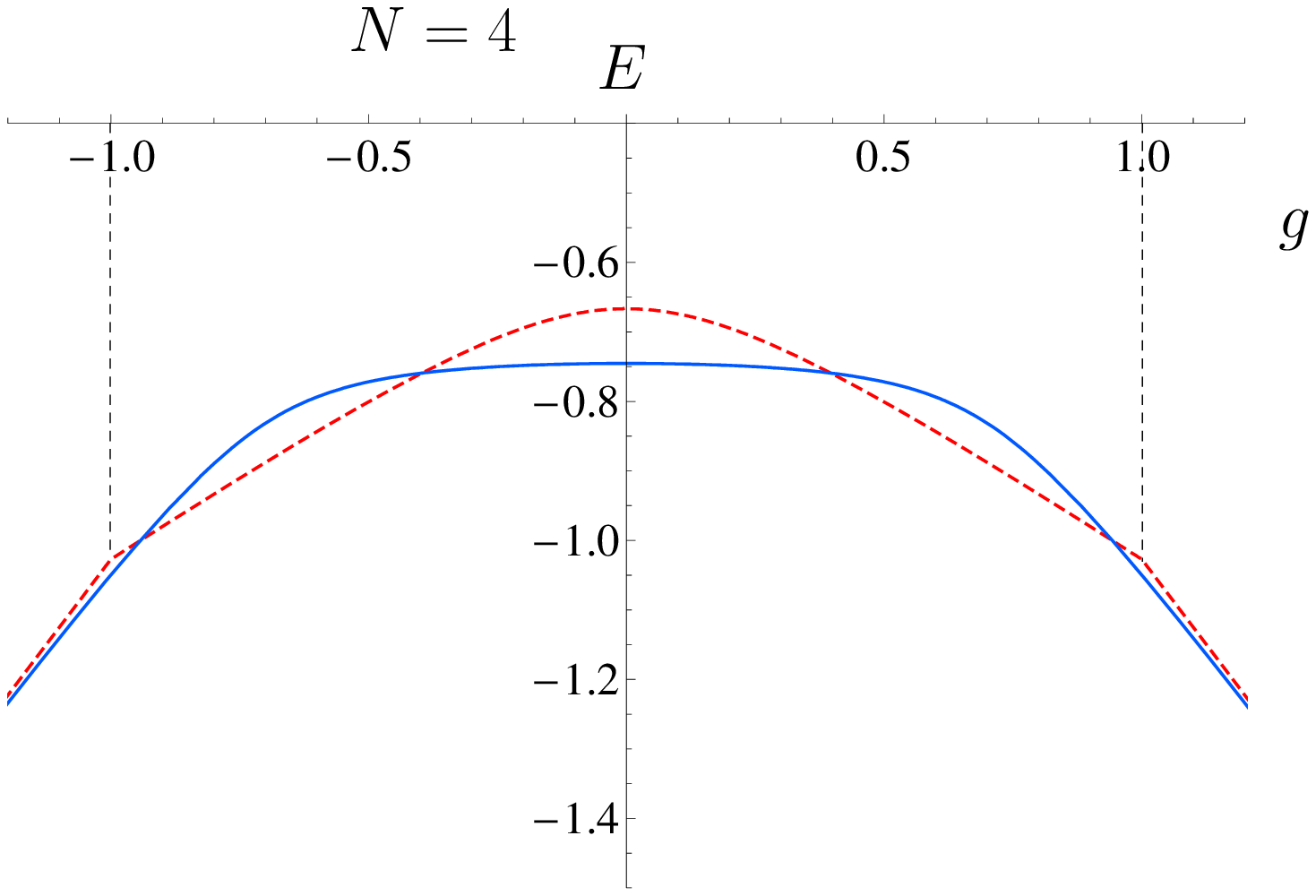}
\quad \quad \\
\includegraphics[width=0.7\columnwidth]{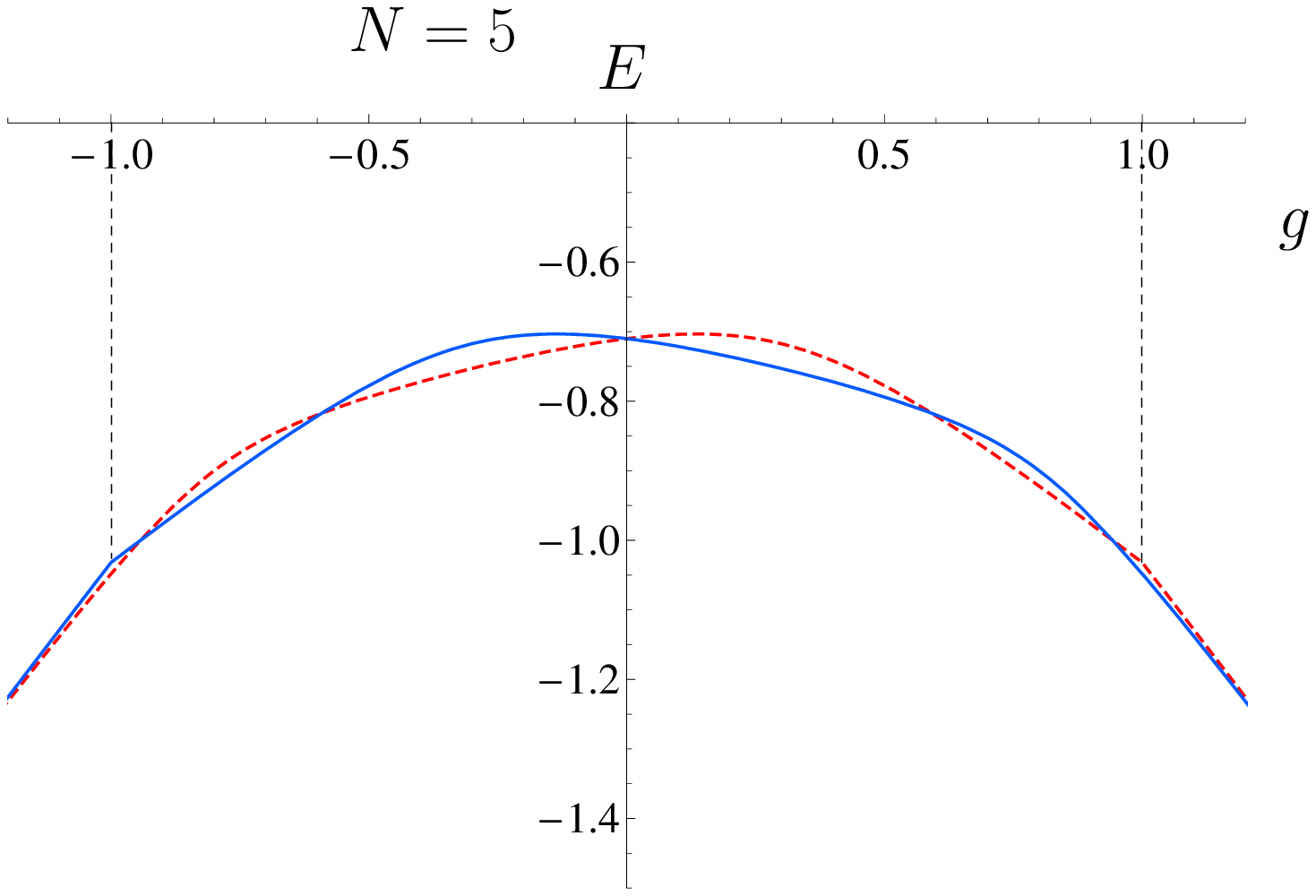}
\caption{Vacua competition for $N=4$ and $5$ spins: for both cases
the dashed line corresponds to $E_{\rm vac}^{(-)}$ and the solid one
to $E_{\rm vac}^{(+)}$.} \label{fig:vacuacompetition}\end{center}
\end{figure}

Therefore, for $g\in(-1,1)$, in the various regions of magnetic
field $g$ the ground state  is alternatively given by  one of the
two vacua with energy (\ref{eq:XYvacuum energy alpha
0})-(\ref{eq:XYvacuum energy alpha 1/2}).  We call this mechanism
\textit{vacua competition} between the two parity sectors. See
Fig.~\ref{fig:vacuacompetition}.

For $g<-1$ the vacuum state with $\varrho=-1$ for $N$ even
($\varrho=+1$ for $N$ odd) is not physical, because it has the wrong
parity $\bar{\varrho}=\varrho=-(-1)^N$, and it is  ruled out from
the competition  by the projection $P_{\bar{\varrho}}$. Analogously,
for $g>1$ the vacuum state with  $\varrho=-1$ for both $N$ even and
odd is ruled out. However, it is not difficult to prove that the energy
of the unphysical vacuum when $|g|>1$ is always larger than the
physical one. Therefore, as far as one is interested in the ground
state, the ground state is the result of the vacua competition in
the \emph{whole} range $g\in\mathbb{R}$. Not so for the first
excited level, which is the energy of the ``losing'' vacuum only in
the range $(-1,1)$, while outside it is the lowest 1-fermion energy
level above the losing vacuum.

\begin{figure}
\begin{center}
\includegraphics[width=0.76\columnwidth]{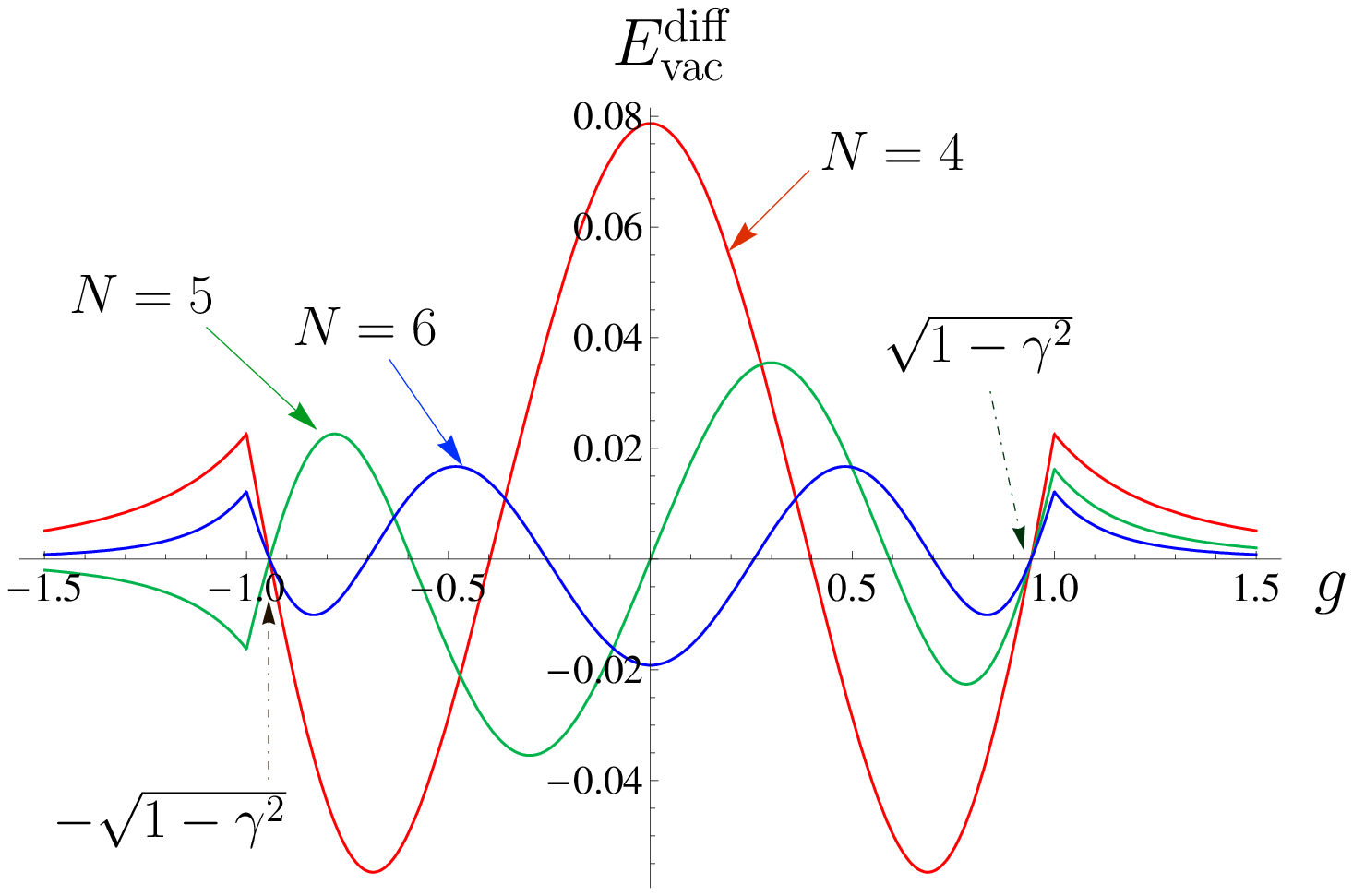}
\caption{Difference between energy densities (in unit $J$) of the two
vacua at $\gamma=\frac{1}{3}$ for even numbers of spins, $N=4, 5, 6$. For all $N$, $E_{\rm vac}^{\rm diff}$
vanishes at $g=\sqrt{1-\gamma^2}$.
} \label{fig:diff tra i vuoti}
\end{center}
\end{figure}
More generally, from (\ref{eq:rhog even}) and (\ref{eq:rhog odd}) it
easily follows that the whole spectrum is given for $g\in(-1,1)$ by
the union of the spectra of eigenstates with an even number of
Bogoliubov fermions ($\bar{\varrho}=(-1)^N$) of both Hamiltonians
$\tilde{H}_\gamma^{(\varrho)}$ with $\varrho=\pm1$. On the other
hand, outside the above interval, the spectrum is given by the
eigenstates of $\tilde{H}_\gamma^{(\varrho)}$
($\tilde{H}_\gamma^{(-\varrho)}$) with an even (odd) number of
Bogoliubov particles, where $\varrho=(-1)^N$ for $g<-1$ and
$\varrho=+1$ for $g>1$. The intersection points between the vacua
energy densities depend in general on the number of spins $N$;
however, independently of $N$,  the difference between the two
energy densities,
\begin{equation}
\label{eq:XYvacua_difference}
\fl E_{\rm vac}^{\rm diff}(g)=E_{\rm vac}^{(-)}-E_{\rm vac}^{(+)}
=-\frac{1}{N}\sum_{m\in
\mathbb{Z}_{2N}}(-1)^m \sqrt{\left[g-\cos\left(\frac{\pi
m}{N}\right)\right]^2+\gamma^2\sin^2\left(\frac{\pi m}{N}\right)},
\end{equation}
always vanishes at
$g=\pm\sqrt{1-\gamma^2}$ (see \fref{fig:diff tra i vuoti}).
Indeed one has
\begin{eqnarray}
E_{\rm vac}^{\rm
diff}(\pm\sqrt{1-\gamma^2})&=&-\frac{1}{N}\sum_{k\in
\mathbb{Z}_{2N}} (-1)^m\left[ 1 \mp\sqrt{1-\gamma^2}
\cos\left(\frac{\pi m}{N}\right)\right]
\end{eqnarray}
and
\begin{equation}
\sum_{k\in \mathbb{Z}_{2N}} (-1)^m \cos\left(\frac{\pi
m}{N}\right)={\rm Re}\left(\frac{1-\mathrm{e}^{\mathrm{i} 2\pi
(N+1)}}{1-\mathrm{e}^{\mathrm{i} \pi (N+1)/N}} \right)=0.
\end{equation}
From \fref{fig:diff tra i vuoti} on can also observe that for finite
size systems the vacua intersection points present discontinuities
of the first derivative, as will be explicitly shown in
\sref{sec:QTP}. In that section we will also focus on the points
$g=\pm1$ which are two interesting values of the magnetic field for
this class of Hamiltonians, since they will be shown to represent
the finite-size forerunners of the quantum phase transition points
(in the thermodynamic limit).

\section{The XX model}\label{sec:theXXmodel}
The XX model ($\gamma=0$) is known as the isotropic model since
the interaction between nearest neighbours spins along $x$ and $y$
axis is characterized by the same coefficient in the Hamiltonian  (\ref{eq:XYhamiltonian}):
\begin{equation}\label{eq:XXhamiltonian}
H_{XX}(g)= H_{\gamma=0}(g)=-J \sum_{i\in\mathbb{Z}_{N}}\Big[ g\sigma_i^z +
\frac{1}{2}\sigma_i^x \sigma_{i+1}^x + \frac{1}{2}\sigma_i^y
\sigma_{i+1}^y \Big].
\end{equation}
In this case equation \eref{eq:XY hamiltonian with deformed
Fourier transform} reduces to
\begin{eqnarray}\label{eq:XX hamiltonian with deformed Fourier transform}
\fl \qquad H_{0}^{(\varrho)}(g)=2 J\sum_{k\in\mathbb{Z}_{N}}
 \left[\cos\left(2\pi\frac{\alpha+k}{N}\right)-g\right]
\left(\hat{c}_k\hat{c}_k^\dagger-\frac{1}{2}\right) P_\varrho
             , \qquad \alpha=\frac{1+\varrho}{4}.
\end{eqnarray}
From this follows that the Fourier transformed XX Hamiltonian is already diagonal  and
the last term characterizing coupled fermions in \Eref{eq:XY
hamiltonian with deformed Fourier transform} vanishes for all $k$.
In other words in the XX model we are only dealing with single
fermions,  $\mathcal{S}_\varrho = \mathbb{Z}_N$, and the Bogoliubov transformation (\ref{eq:total Bogoliubov}) reduces to
\begin{equation}\label{eq:Bogoliubov for XX model}
U_{B}(g;\varrho,\bm{s})=\prod_{k \in \mathbb{Z}_N} C_k^{s_k},
\end{equation}
where now $\bm{s} \in \{0,1\}^N$ is an \emph{unconstrained} binary string of length $N$. This yields
\begin{eqnarray}
\tilde H_0^{(\varrho)}(g) &=& U_{B}(g;\varrho,\bm{s}) H_0^{(\varrho)}(g) U_{B}(g;\varrho,\bm{s})^\dagger
\nonumber\\
&=& 2 J\sum_{k\in\mathbb{Z}_{N}}
 (-1)^{s_k} \left[\cos\left(2\pi\frac{\alpha+k}{N}\right)-g\right]
\left(\hat{c}_k\hat{c}_k^\dagger-\frac{1}{2}\right) P_{\bar{\varrho}}
\label{eq:HXX Bog}
\end{eqnarray}
with $\bar{\varrho}= (-1)^{|\bm{s}|} \varrho$.
 In particular, if $s_k=0$ the Bogoliubov transformation associates JW fermions to
Bogoliubov fermions, while if $s_k=1$ it transforms JW fermions into
Bogoliubov antifermions, or holes.
\subsection{The energy spectrum}
\begin{figure}
\begin{center}
\includegraphics[width=0.45\columnwidth]{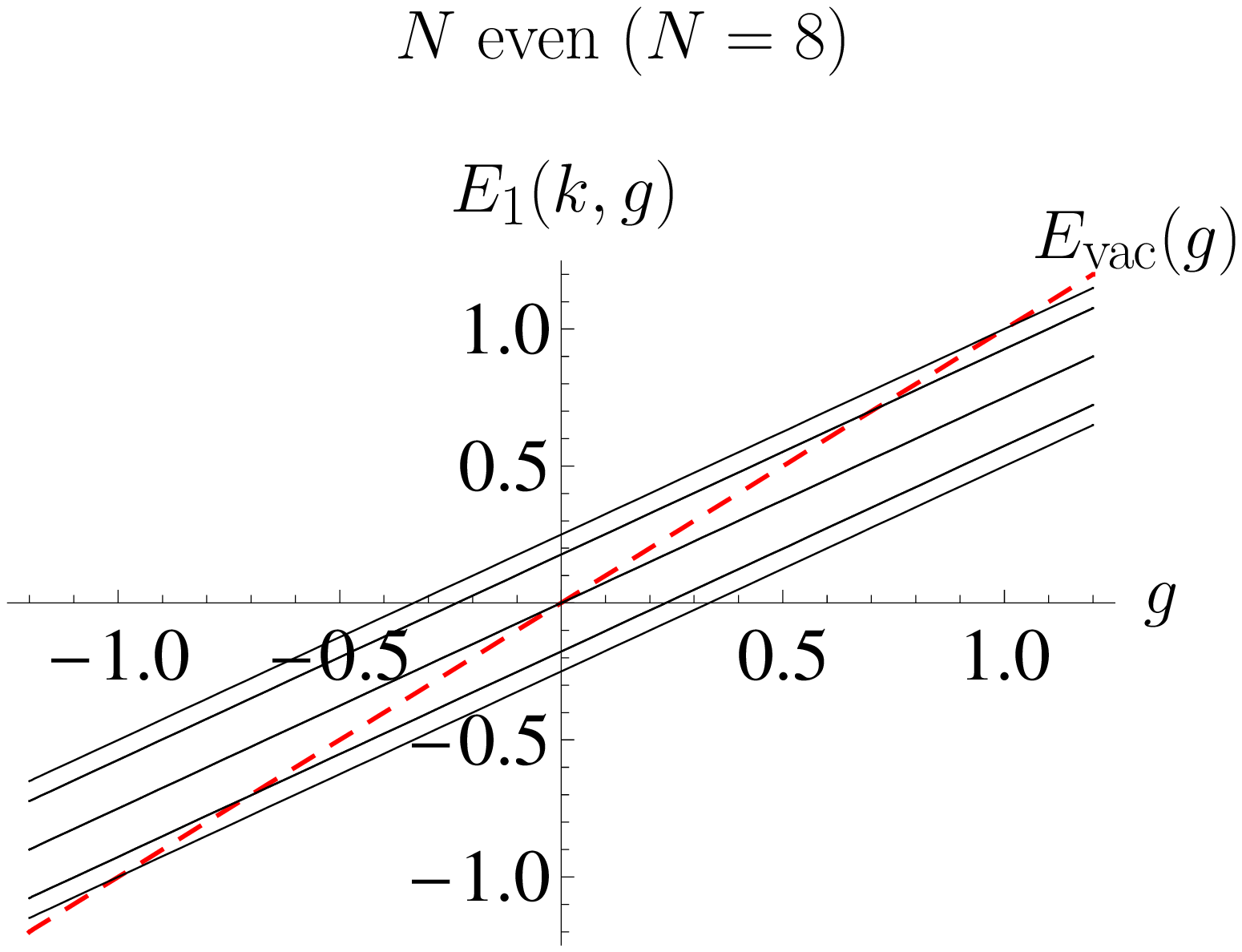}
\qquad
\includegraphics[width=0.45\columnwidth]{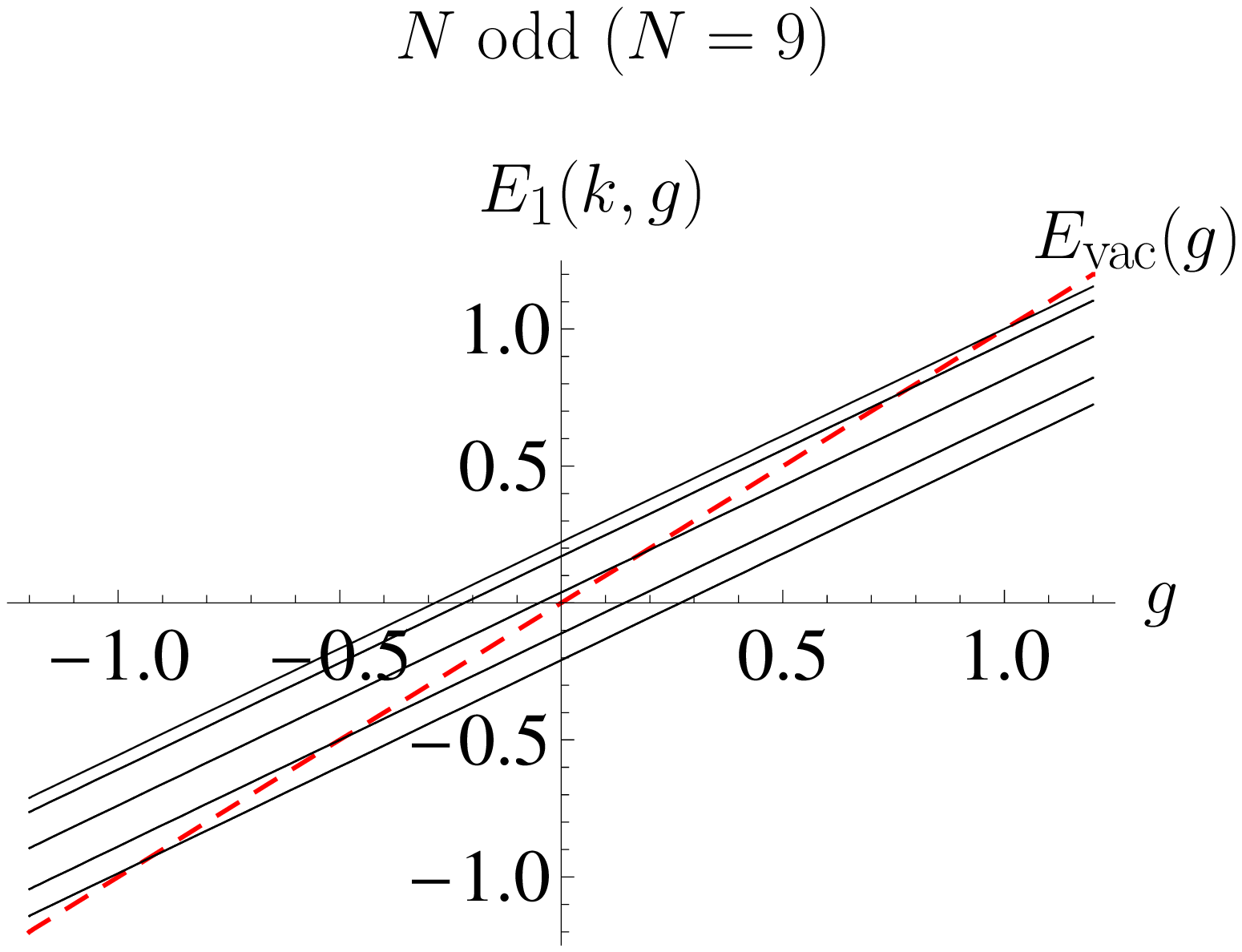}
\end{center}
\caption{Single particle energy densities $E_1(k,g)$ (solid
lines) and vacuum energy densities $E_{\rm vac}(g)$
(dashed lines). Different lines correspond to different
$k\in\mathbb{Z}_N$ according to \eref{eq:XXsingleenergy}.}
\label{fig:XXsingle particle}
\end{figure}

\begin{figure}
\begin{center}
\begin{tabular}{cc}
\includegraphics[width=0.47\columnwidth]{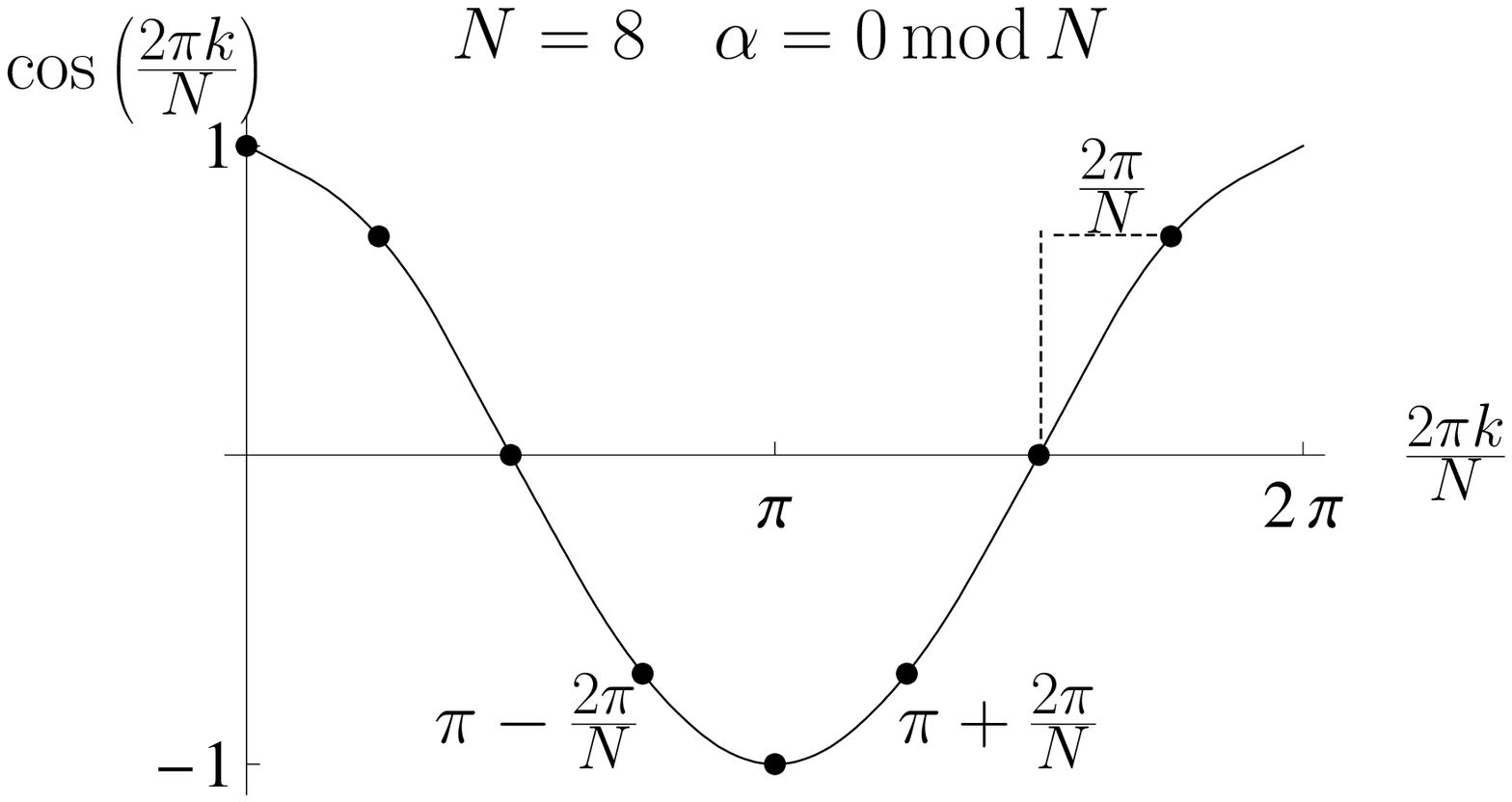}
&
\includegraphics[width=0.47\columnwidth]{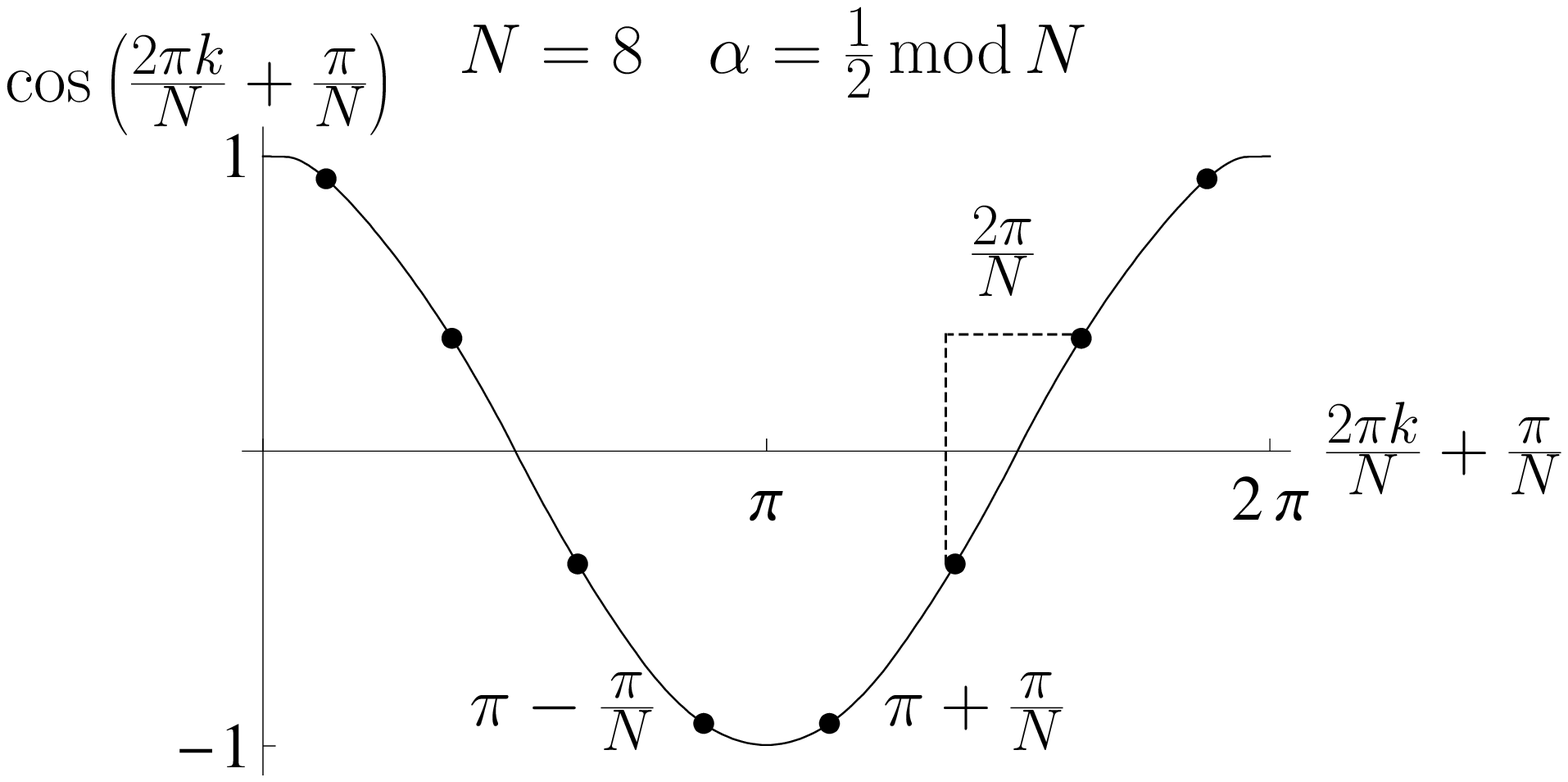}\\[5mm]
\includegraphics[width=0.47\columnwidth]{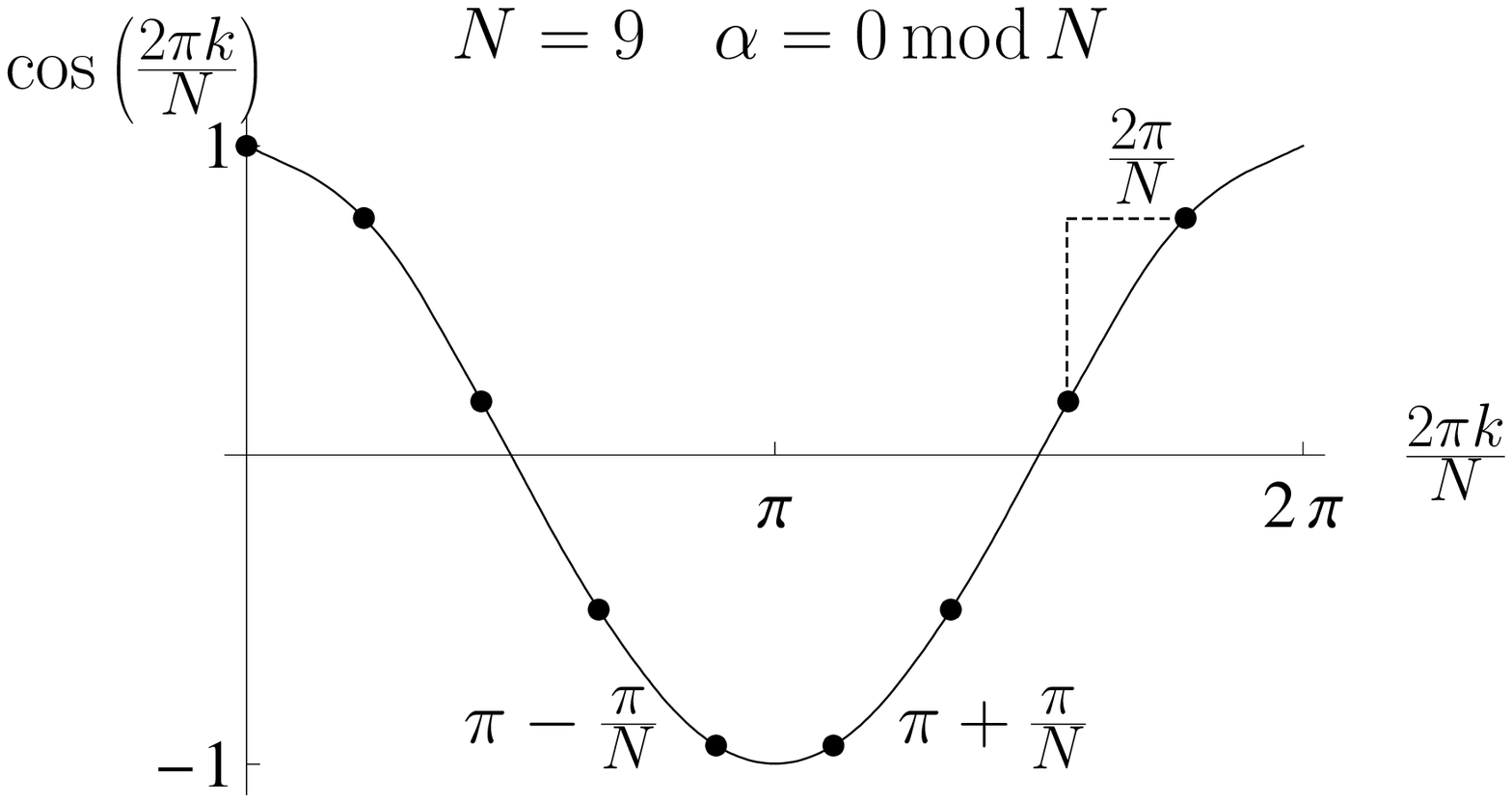}
&
\includegraphics[width=0.47\columnwidth]{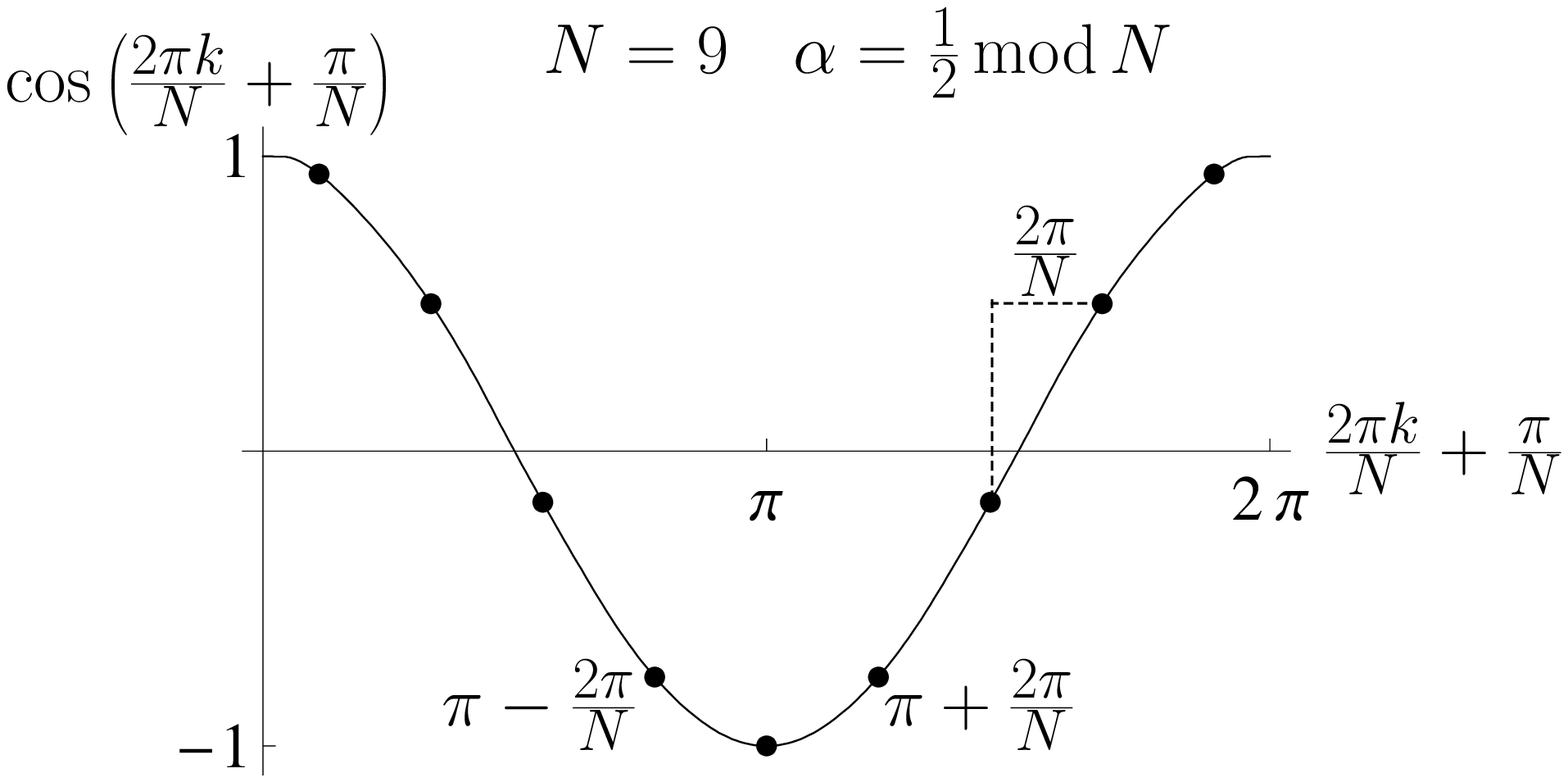}
\end{tabular}
\end{center}
\caption{Plot of $\cos[2\pi(\alpha+k)/N]$, $k\in\mathbb{Z}_N$ for
$N=8,9$ and $\alpha\equiv 0, 1/2\; (\mod N)$.}
\label{fig:XXcos}
\end{figure}

\begin{figure}
\begin{center}
\includegraphics[width=\columnwidth]{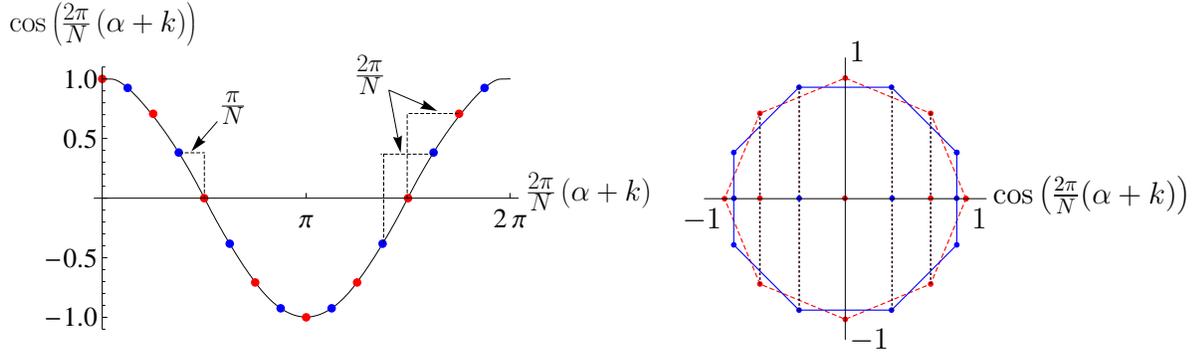}
\end{center}
\caption{(Color online) Left: plot of $\cos[2\pi(\alpha+k)/N]$,
$k\in\mathbb{Z}_N$ with $\alpha\equiv 0\; (\mod N)$ (dashed line) and
$\alpha=1/2\; (\mod N)$ (solid line), for $N=8$. Right:
geometrical description of $\cos[2\pi(\alpha+k)/N]$ for $N=8$.}
\label{fig:NfixedCompares}
\end{figure}

As already emphasized at the end of section \ref{sec:the bogoliubov
transform}, the energy spectrum does not depend on the choice of the
gauge $\bm{s}$ of the unitary Bogoliubov transformation. If
$\bm{s}=0$ equation \eref{eq:HXX Bog} becomes
\begin{equation}\label{eq:XXhamDiag}
\tilde H_0^{(\varrho)}= H_0^{(\varrho)} = 2 J\sum_{k\in\mathbb{Z}_{N}}\left[\cos \left(
2\pi\frac{\alpha+k}{N} \right)-g\right] \left(\hat{c}_k^\dag
\hat{c}_k-\frac{1}{2}\right) P_{\varrho},
\end{equation}
The spectrum of the above Hamiltonian, and in particular its ground
state energy has been studied in \cite{XX Model}. We quickly
summarize the main results and show how they derive from vacua
competition. The energy density$/J$  of the vacuum state does not
depend on the parity $\varrho$ and on the size $N$
\begin{equation}\label{eq:XXenervacuum}
E_{\rm{vac}}(g) =\frac{1}{N}\sum_{k\in\mathbb{Z}_{N}}\left[
g-\cos\!\left(2\pi \frac{\alpha+k}{N}\right) \right]=g.
\end{equation}
On the other hand, if we add one fermion of momentum $k$ the energy density reads
\begin{equation}\label{eq:XXsingleenergy}
E_1^{(\varrho)}(k,g)=E_{\rm {vac}}(g)-\frac{2}{N}\left[
g-\cos\!\left(2\pi\frac{\alpha+k}{N}\right) \right],
\qquad \alpha=\frac{1+\varrho}{2}
\end{equation}
where $\varrho=-(-1)^N$ is the parity of the 1-particle sector.
In \fref{fig:XXsingle particle} we represent the single particle
energy spectra corresponding to $N=8$ and $9$ sites
(representative of an even/odd number of spins,
respectively). The different lines are parametrized by
$k\in\mathbb{Z}_N$ and one notes the presence of degeneracies in
both cases. Since we are interested in the ground state of the
system, we focus on the lowest energy levels and consider the
values assumed by the function $\cos[2\pi(\alpha+k)/N]$ in the
four possible cases
($N$ even or odd and $\alpha\equiv 0$ or $1/2 \mod N$), as shown
in Fig. \ref{fig:XXcos}. Notice that these results can be
described in terms of regular polygons inscribed in a circle of
unit radius, see \fref{fig:NfixedCompares}.

\begin{figure}
\begin{center}
\includegraphics[width=0.8\columnwidth]{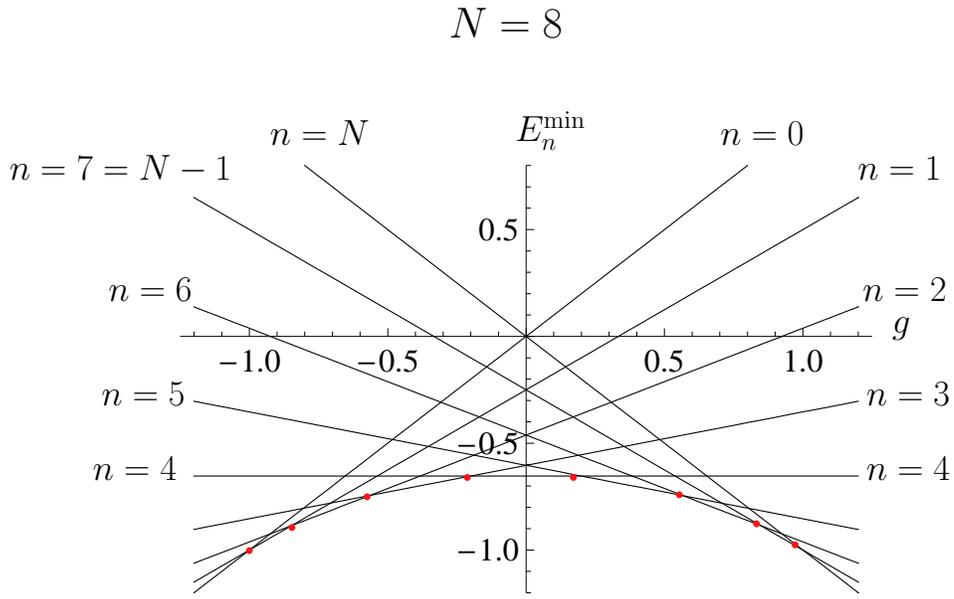}
\end{center}
\caption{Lowest energy levels $E_n^{\rm min}(g)$ for
different number of fermions $n$; the intersection between the
energy levels corresponding to $n$ and $n+1$ fermions (starting
from $n=0$) are the points of level crossing
(dots).}\label{fig:Points of level crossing}
\end{figure}

From  \fref{fig:XXcos} one obtains the values of $k$ that minimize
the energy per site; in the 1-particle sector one has
\begin{equation}\label{eq:1partMomentum}
\left\{
\begin{array}{l@{\quad}l@{\quad}l@{}l}
\medskip
N\ {\rm even }&\Rightarrow&
\alpha\equiv0&\ {\rm mod}\ N\Rightarrow k=\left[\frac{N}{2}\right],\\
N\;{\rm odd}&\Rightarrow& \alpha\equiv\frac{1}{2}&\ {\rm mod}\
N\Rightarrow k=\left[\frac{N-1}{2}\right].
\end{array}
\right.
\end{equation}
Similarly, in the 2-particle sector the energy is  minimum for
\begin{equation}
\left\{
\begin{array}{l@{\quad}l@{\quad}l@{}l}
\medskip
N\;\mbox{even}&\Rightarrow&\alpha \equiv \frac{1}{2}&\ {\rm mod}\
N\Rightarrow
\{k_1,k_2\}=\left\{\left[\frac{N}{2}-1\right],\left[\frac{N}{2}\right]\right\}, \\
N\;\mbox{odd}&\Rightarrow&\alpha\equiv0&\ {\rm mod} \ N\Rightarrow
\{k_1,k_2\}=\left\{\left[\frac{N-1}{2}\right],\left[\frac{N+1}{2}\right]\right\}.
\end{array}
\right.
\end{equation}
It turns out that the general expression of the lowest energy levels in
the different $n$-particle sectors  does not depend on the parity
of $N$. For $n$ fermions, one gets \cite{XX Model}
\begin{equation}\label{eq:lowestEnergyLevels}
E_n^{\rm min}(g)=g \left(1-\frac{2n}{N}\right) -
\frac{2}{N}\frac{\sin(n\pi/N)}{\sin(\pi/N)}.
\end{equation}
In \fref{fig:Points of level crossing} we plot the lowest energy
levels corresponding to $0\leq n \leq N$ for $N=8$ sites. The
intersections of levels corresponding to $n$ and $n+1$ fermions
(starting from $n=0$) define the \textit{points of level crossing}
$g_{c}$, where an excited level and the ground state are
interchanged. The analytic expression of the critical points is
easily obtained by the condition $E_n^{\rm min}(g_{c})=E_{n+1}^{\rm min}(g_{c})$. We find
\begin{eqnarray}
\label{eq:XXlevelcrossingPoints}
g_{c}(n)&=&(-1)^{n+1}
\left[1+2\sum_{m=1}^{n}\cos\left(\frac{\pi
m}{N}\right)\right]\\&=&\frac{
\sin(n\pi/N)-\sin[(n+1)\pi/N]}{\sin(\pi/N)}, \qquad 0\leq n\leq
N-1.
\end{eqnarray}
As a consequence, the ground-state energy density is
\begin{equation}\label{eq:enermin}
\fl \qquad E_{\rm gs}(g)=g \left(1-\frac{2n}{N}\right) -
\frac{2}{N}\frac{\sin(n\pi/N)}{\sin(\pi/N)}\quad {\rm with}\quad
g\in \Big( g_{c}(n-1),g_{c}(n)\Big),
\end{equation}
with $0\leq n\leq N$ and where we stipulated that $g_{c}(-1)=-\infty$ and
$g_{c}(N)=+\infty$. Thus, for $g\in (g_c(n-1),g_{c}(n))$, the
ground state contains $n$ JW fermions. Note that $g_{c}(0)=-1$ and
$g_{c}(N-1)=+1$, independently of $N$.

\begin{figure}
\begin{center}
\includegraphics[width=0.8\columnwidth]{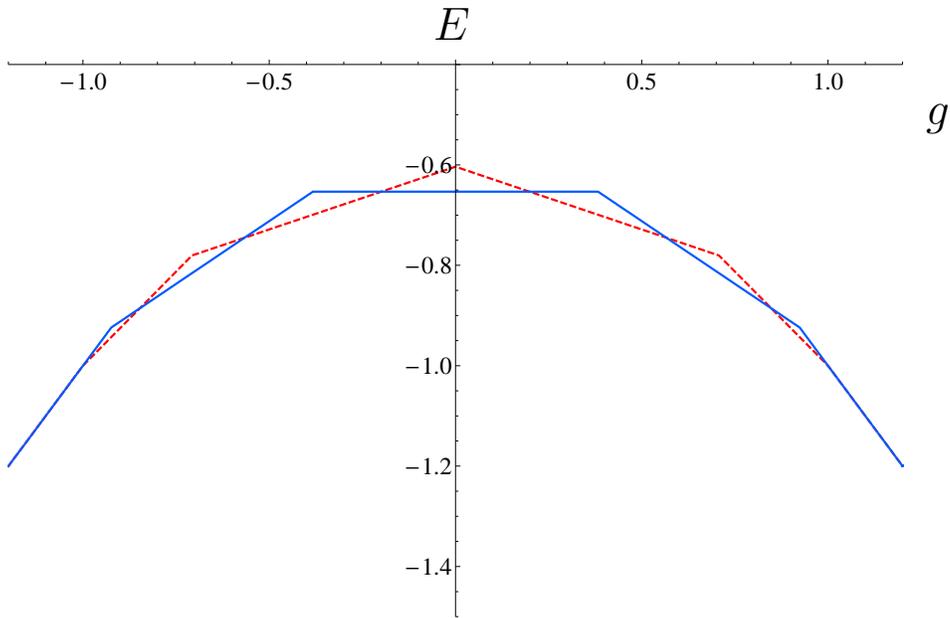}
\end{center}
\caption{The ground state energy density of the XX model ($N=8$) is given by the
competition between the vacua energy densities $E_{\rm vac}^{(-)}$
(dashed line) and $E_{\rm vac}^{(+)}$ (solid
line).}\label{fig:XXvacua_competition}
\end{figure}

We will now derive the ground state energy density  starting from
the same choice of the Bogoliubov transform made for the XY
Model (\ref{eq:H vacua comp}) that particularizes to
\begin{equation}
\tilde{H}^{(\varrho)}_0(g) =2 J \sum_{k\in\mathbb{Z}_{N}}
\left|\cos \left(
2\pi\frac{\alpha+k}{N} \right)-g\right|
\left(\hat{c}_k^\dag
\hat{c}_k-\frac{1}{2}\right) P_{\bar{\varrho}(g)},
\end{equation}
\begin{equation}
\bar{\varrho}(g)=(-1)^{|\bm{s}(g)|}\varrho.
\end{equation}
The ground state is then the winner of  the vacua competition
between \numparts
\begin{eqnarray}
E_{\rm vac}^{(-)}(g)=-\frac{1}{N}\sum_{k\in\mathbb{Z}_{N}}
\left|\cos\left( \frac{2\pi k}{N}
\right)-g\right|\label{eq:XXvacuum_alpha0},\\
E_{\rm vac}^{(+)}(g)=-\frac{1}{N}\sum_{k\in
\mathbb{Z}_{N}} \left|\cos\left( \frac{2\pi k}{N}+\frac{\pi}{N}
\right)-g\right|\label{eq:XXvacuum_alpha1/2}
\end{eqnarray}
\endnumparts
(see Fig. \ref{fig:XXvacua_competition}).
The  points of level crossing (\ref{eq:XXlevelcrossingPoints}) are given
by those values of the magnetic field that satisfy the following
equation
\begin{equation}\label{eq:diff}
E^{\rm diff}_{\rm vac}(g)=-\frac{1}{N}\sum_{k\in\mathbb{Z}_{2N}}(-1)^k
\left|\cos\left( \frac{\pi k}{N} \right)-g\right|=0.
\end{equation}

\begin{figure}
\begin{center}
\includegraphics[width=0.6\columnwidth]{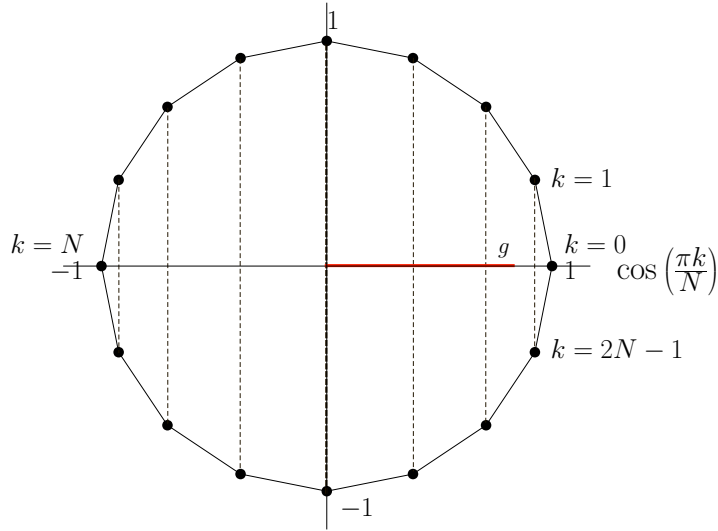}
\end{center}\caption{Geometrical representation of \Eref{eq:diff} when $N=8$; the thick red line is the magnetic
field $g$. \label{fig:poligonoxx}}
\end{figure}
Consider the regular polygon inscribed in a circle of unit radius in
\fref{fig:poligonoxx}; it is a geometrical representation of the
function $\cos\left( \frac{\pi k}{N} \right)$ for $k \in
\mathbb{Z}_N$. When $|g|>1$ one immediately gets $E^{\rm diff}_{\rm
vac}(g)=0$, whereas for $|g|\leq1$ the key idea is to consider the
$N$ intervals on the $x$ axis limited by the dashed vertical lines,
represented in \fref{fig:poligonoxx}; for each interval one can
write the explicit expression for the vacua difference
(\ref{eq:diff}). For example when $g \in
\left[\cos\left(\frac{\pi}{N}\right),1\right]$ one gets
\begin{eqnarray}
E^{\rm diff}_{\rm vac}(g)&=&-\frac{1}{N}\left[(1-g)+\sum_{k\in
\mathbb{Z}_{2N}, k \neq [0]}(-1)^k
\left(g-\cos\left( \frac{\pi k}{N} \right)\right)\right]\nonumber\\
&=&-\frac{1}{N}\left[2(1-g)+\sum_{k\in
\mathbb{Z}_{2N}}(-1)^k
\left(g-\cos\left( \frac{\pi k}{N} \right)\right)\right]
\nonumber\\
&=& -\frac{2}{N}(1-g),
\label{eq:diff1}
\end{eqnarray}
from which follows that in this interval $E^{\rm diff}_{\rm
vac}(g)=0$ for $g=1$. Similarly when $g \in \left[\cos
\frac{(m+1)\pi }{N}, \cos \frac{m\pi}{N} \right] $, for $0\leq m
\leq N-1$, one gets that $E^{\rm diff}_{\rm vac}(g)=0$ when
\begin{equation}
g=(-1)^m\left(1+2\sum_{k=1}^m 
\cos\left(\frac{\pi
k}{N}\right)\right)=-g_{c}(m),\qquad 0\leq m\leq N-1,
\end{equation}
where $g_{c}(m)$ are the points of level crossing
\eref{eq:XXlevelcrossingPoints} [for $n=0$ one gets $g=1$, in
agreement with \eref{eq:diff}]. By considering the symmetry
$g_{c}(n)=-g_{c}(N-1-n)$, one immediately sees that the level
crossing points have the same analytic expression of the
intersection points between the two vacua, for $n=N-1-m$.
\section{Thermodynamic  limit and  quantum phase
transitions}\label{sec:QTP}
\subsection{Quantum phase transitions
in the XY model}\label{sec:XY Model QPTs} In this section we will
show that in finite size systems one can find the forerunners of the
points of quantum phase transition. These points are characterized
by the presence of large values of the second derivative of the ground state energy  density,
 that is then amplified and becomes a singularity in
the thermodynamic limit.

\begin{figure}
\begin{center}
\includegraphics[width=0.6\columnwidth]{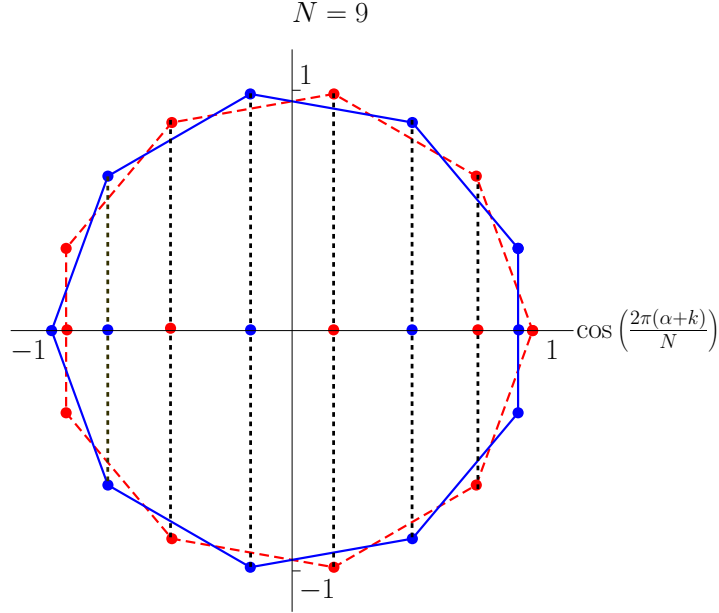}
\end{center}
\caption{Geometrical description for $\cos\left(\frac{2 \pi
(\alpha+k)}{N}\right)$ for $N=9$, similar to
\fref{fig:NfixedCompares}: the two polygons belong to the parity
sectors $\alpha=0\; (\mod N)$ (dashed line) and $\alpha\equiv 1/2\; (\mod N)$
(solid line) } \label{fig:XYpolygonNodd}
\end{figure}

As observed in \sref{sec:XY vacua competition}, the first derivative
of the ground state energy evaluated at the intersection points
between the two vacua is not continuous and for finite size systems
the second derivatives diverges at these points; however we will
show that these  singularities vanish when $N\rightarrow \infty$.
Consider for example the level crossing at $g=\sqrt{1-\gamma^2}$; the difference
between the first derivatives of the two vacuum energies is given
by the derivative of (\ref{eq:XYvacua_difference}):
\begin{eqnarray}\label{eq:energy_difference_first_derivative}
\frac{{\rm d}E_{\rm vac}^{\rm
diff}}{{\rm
d}g}(\sqrt{1-\gamma^2})&=&\frac{1}{N}\frac{\gamma^2}{\sqrt{1-\gamma^2}}\sum_{k\in
\mathbb{Z}_{N}}
\Bigg[\frac{1}{1-\sqrt{1-\gamma^2}\cos\left({\frac{2\pi
k}{N}+\frac{\pi}{N}}\right)}\nonumber\\
&&
\phantom{\frac{1}{N}\frac{\gamma^2}{\sqrt{1-\gamma^2}}\sum_{k\in
\mathbb{Z}_{N}}
\Bigg[}
-\frac{1}{1-\sqrt{1-\gamma^2}\cos\left({\frac{2\pi
k}{N}}\right)}\Bigg].
\end{eqnarray}
When the number of spins $N$ is odd, for each $k \in \mathbb{Z}_{N}$
there is a given $\tilde{k}=k+\frac{N}{2}$ such that
$\cos\left(\frac{2\pi k}{N}\right)=-\cos\left(\frac{2\pi
\tilde{k}}{N}+\frac{\pi}{N}\right)$ (see Fig.\  \ref{fig:XYpolygonNodd}), and the last equation becomes
\begin{eqnarray}
\fl \;\; \frac{{\rm d}E_{\rm vac}^{\rm diff}}{{\rm
d}g}\left(\sqrt{1-\gamma^2}\right)=\frac{1}{N}\frac{\gamma^2}{\sqrt{1-\gamma^2}}\left[
\frac{2 \sqrt{1-\gamma^2}}{\gamma^2}+ 4
\sum_{k=1}^{(N-1)/2}\frac{\cos\left(\frac{2 \pi k}{N}\right)
\sqrt{1-\gamma^2}}{1-\cos\left(\frac{2\pi k}{N}\right)
(1-\gamma^2)}\right];
\end{eqnarray}
from the symmetries of the function $\cos\left(\frac{2 \pi
k}{N}\right)$ one gets that the last expression is strictly greater
than zero. From this it follows that the second derivative of the
vacua energy difference diverges for all finite $N$ at
$g=\sqrt{1-\gamma^2}$, and the same argument can be  extended
to all intersection points between the two vacua.
The case $N$ even, see \fref{fig:NfixedCompares}, is analogous, as
one can see by noting that the polygon corresponding to
$\alpha\equiv\frac{1}{2} (\mod N)$ is rotated by an angle
$\frac{\pi}{N}$ (or in other words, it associates to each momentum
$k\mapsto \tilde{k}=k+\frac{N-1}{2}$).

Summarizing, for finite size systems the second derivative of the
energy density of the
 ground state diverges at the intersection points of the two vacua; on the other
hand in the thermodynamic limit this divergence is suppressed.
Indeed, in the limit $N\to\infty$ equation
\eref{eq:energy_difference_first_derivative} becomes:
\begin{equation}\label{eq:diff_firt_derivative_in_thermod_limit}
\frac{{\rm d}E_{\rm vac}^{\rm diff}}{{\rm
d}g}\left(\sqrt{1-\gamma^2}\right)\sim \frac{\gamma^2}{2 \pi
\sqrt{1-\gamma^2}}\int_{0}^{2\pi} {\rm d}x
\left(f(x)-f(x+\pi/N)\right),
\end{equation}
where $f(x)$ is given by:
\begin{equation}
f(x)=\frac{1}{1-\sqrt{1-\gamma^2}\cos(x)}.
\end{equation}
Expanding in Taylor series $f(x+\pi/N)$ 
one gets
\begin{equation}
\frac{{\rm d}E_{\rm vac}^{\rm diff}}{{\rm
d}g}\left(\sqrt{1-\gamma^2}\right)\to \frac{\gamma^2}{\pi
\sqrt{1-\gamma^2}}f'(0)=0, \qquad N\to\infty.
\end{equation}
This means that the singularities of the second derivative of the
ground state vanish in the thermodynamic limit; in other words,
the forerunners of the quantum phase transition are \textit{not}
related to finite-size level crossings of the ground state. In this
section we will show that they are related to the level crossings
between the unphysical vacuum and the losing physical vacuum
where single Bogoliubov fermions sit.

Consider the explicit expressions of the
vacua energies corresponding to the four possible cases given by
 the parity of $N$  and the two parity sectors
\numparts
\begin{enumerate}
\item 
$N$ even, $\varrho=-1$, $\mathcal{S}_\varrho =\left\{[0], \left[\frac{N}{2}\right]\right\}$,
\begin{eqnarray}\label{eq:vacuum_alpha0Neven}
\fl E_{\rm vac}^{(-)}=-\frac{1}{N}\Bigg[\sum_{k\in\mathcal{C}_\varrho}
\sqrt{\left[g-\cos\left(\frac{2\pi
k}{N}\right)\right]^2+ \gamma^2 \sin^2\left(\frac{2\pi
k}{N}\right)}+ |g-1|+|g+1|\Bigg];
\end{eqnarray}
\item 
$N$ odd, $\varrho=-1$, $\mathcal{S}_\varrho =\left\{[0]\right\}$,
\begin{eqnarray}
\fl E_{\rm vac}^{(-)}=-\frac{1}{N}\Bigg[\sum_{k\in\mathcal{C}_\varrho}
\sqrt{\left[g-\cos\left(\frac{2\pi
k}{N}\right)\right]^2+ \gamma^2 \sin^2\left(\frac{2\pi
k}{N}\right)}+ |g-1|\Bigg];
\end{eqnarray}
\item 
$N$ even, $\varrho=+1$, $\mathcal{S}_\varrho =\emptyset$,
\begin{eqnarray}
\fl E_{\rm vac}^{(+)}=-\frac{1}{N}\sum_{k\in\mathbb{Z}_N}
\sqrt{\left[g-\cos\left(\frac{2\pi
k}{N} + \frac{\pi}{N}\right)\right]^2+ \gamma^2 \sin^2\left(\frac{2\pi
k}{N}+ \frac{\pi}{N}\right)};
\end{eqnarray}
\item 
$N$ odd, $\varrho=+1$, $\mathcal{S}_\varrho =\left\{\left[\frac{N-1}{2}\right]\right\}$,
\begin{eqnarray}
\fl E_{\rm vac}^{(+)}=-\frac{1}{N}\Bigg[\sum_{k\in\mathcal{C}_\varrho}
\sqrt{\left[g-\cos\left(\frac{2\pi
k}{N}+ \frac{\pi}{N}\right)\right]^2+ \gamma^2 \sin^2\left(\frac{2\pi
k}{N}+ \frac{\pi}{N}\right)}+ |g+1|\Bigg].
\nonumber\\
\label{eq:vacuum_alpha1/2Nodd}
\end{eqnarray}
\end{enumerate}
\endnumparts
Observe that the absolute values in the previous expressions
correspond to the cosines evaluated at  single fermion momenta
$\mathcal{S}_\varrho$; at these values of the magnetic field the
 first derivative of energy is not continuous
(see \fref{fig:vacuacompetition}) and 
the second derivative has terms proportional to the Dirac delta functions
$\delta(g\pm1)$. However, remember that the vacuum in case (i)
becomes unphysical as soon as $|g|>1$, so that at $g=\pm 1$ there is
a level crossings between physical and unphysical states. The same
phenomenon happens to the vacuum in case (ii)  at $g=1$, and to the
vacuum in case (iv)  at $g=-1$. On the other hand, one can observe
that for finite size chains, for both even and odd $N$, the ground
state is smooth at $g=\pm1$, in other words the ground state, which
coincides with  the winning vacuum state, does not have  any
singularities at these points. However, it can be shown that the
second derivative of the ground state energy at $g=\pm1$ scales as~$-\log N$ .

\begin{figure}
\begin{center}
\includegraphics[width=0.8\columnwidth]{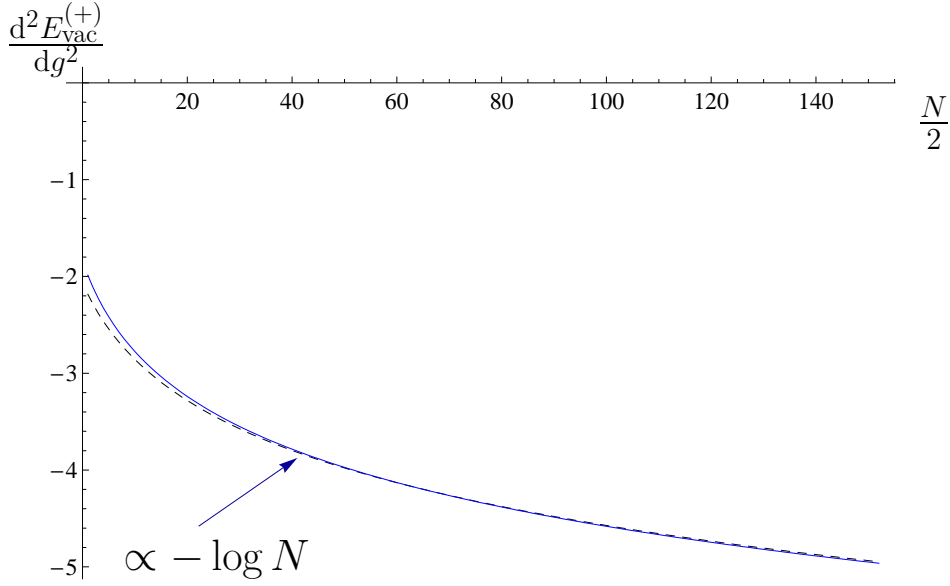}
\end{center}\caption{Second derivative of the vacuum energy density at $g=1$ in the parity sector
$\varrho=+1$  for $N$ even, from $18$ to $320$ (dashed line): it
scales as $-\log N$ (solid line). }\label{fig:LogN}
\end{figure}

Consider for example the case of an even number of spins $N$. In
this case the ground
state belongs to the parity sector with 
$\varrho=+1$, without singularities.
\Fref{fig:secodDerivativeForN=6,24,54} displays ${\rm d}^2 E_{\rm
vac}^{(+)}/{\rm d}g^2$ for $N=6, 24, 54$; at $g=\pm1$ it scales like
$-\log N$.
\begin{figure}
\begin{center}
\includegraphics[width=0.8\columnwidth]{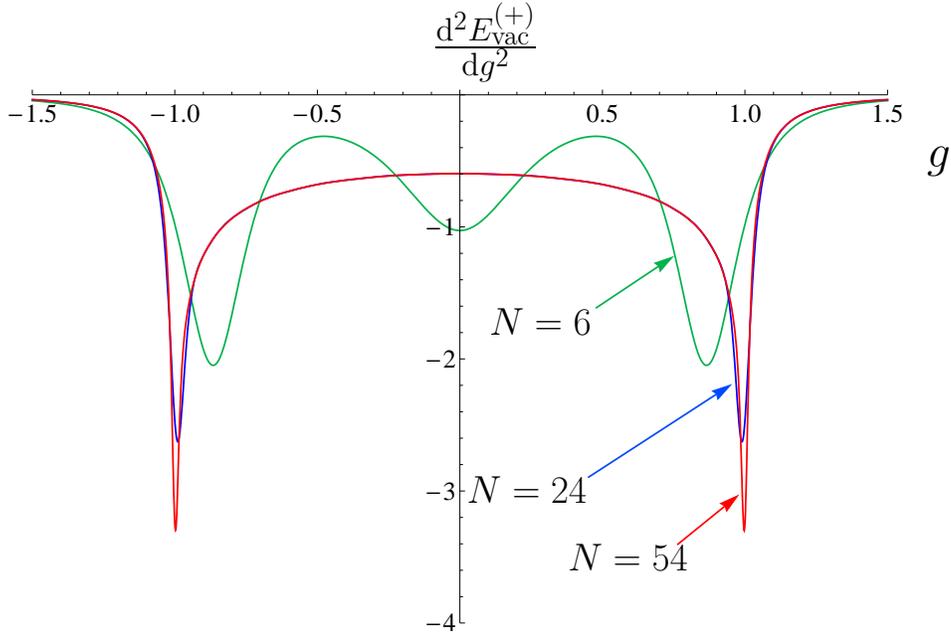}\caption{Second derivative of the vacuum energy density
for even values of  $N$, in the parity sector
$\varrho=+1$.}
\label{fig:secodDerivativeForN=6,24,54}
\end{center}
\end{figure}
Indeed when $g=1$, by deriving (\ref{eq:XYvacuum energy alpha 1/2})
one has
\begin{eqnarray}
\fl \qquad \frac{{\rm d}^2E_{\rm vac}^{(+)}}{{\rm
d}^2g}\left(1\right) &=& -\frac{\gamma^2}{N}\sum_{k\in
\mathbb{Z}_{N}}\frac{\left(1+\cos\left(\frac{2\pi
k}{N}+\frac{\pi}{N}\right)\right)^{3/2}}{\left|\sin\left(\frac{2\pi
k}{N}+\frac{\pi}{N}\right)\right|\left[1+\gamma^2+\cos\left(\frac{2\pi
k}{N}+\frac{\pi}{N}\right)(\gamma^2-1)\right]^{3/2}}\nonumber\\
&\simeq& -\frac{1}{\gamma
\pi}\left[3+\log\left(\frac{N}{8}-\frac{1}{2}\right)\right] -\frac{1}{2}\frac{\gamma^2}{\left(1+\gamma^2\right)^{3/2}}\sim
-\frac{1}{\gamma \pi} \log N,
\end{eqnarray}
for $N\to\infty$, as shown in \fref{fig:LogN}.
The cases $g=-1$ and $N$ even ($\varrho=+1$)
and $g=\pm1$ and $N$ odd 
($\varrho=\mp 1$) are analogous.

\begin{figure}
\begin{center}
\includegraphics[width=0.8\columnwidth]{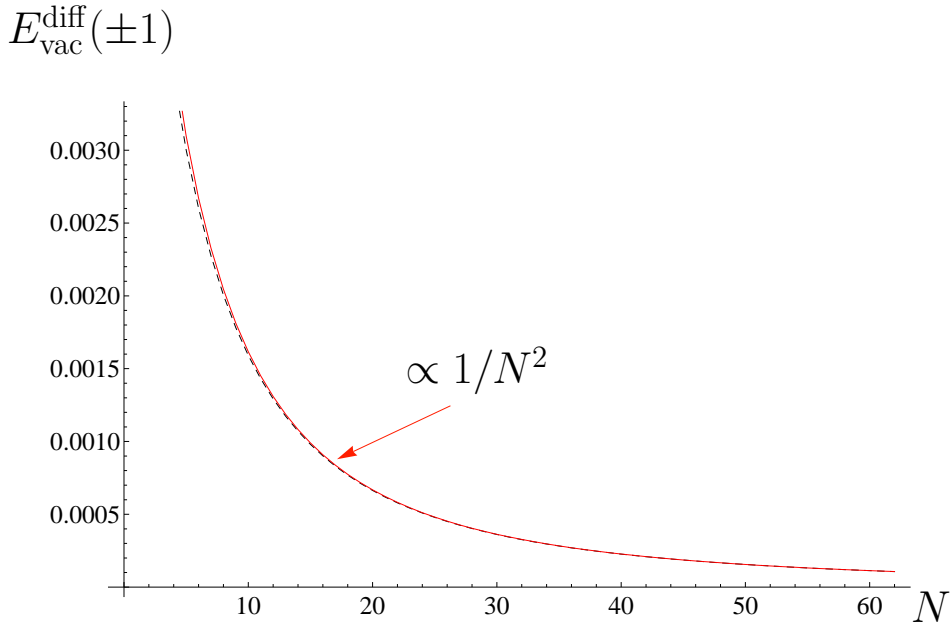}
\caption{Difference between the two vacua energy densities at $g=\pm1$: exact result (dotted line) and asymptotic
approximation of order $1/N^2$ (solid line).}\label{fig:vacua_diff_g=pm1}
\end{center}
\end{figure}

The quantum phase transition is forerun by the losing vacuum whose second
derivative contains a Dirac delta function, at the transition
between physical and unphysical states.
When $N$ tends
to infinity, as we will now show, the difference between the two vacua at $g=\pm 1$
tends to zero and quantum phase transition forerunners approach the
ground state, building up  singularities at logarithmic rates. Indeed, at $g=\pm1$ from
\Eref{eq:XYvacua_difference} one has:
\begin{equation}\label{eq:XYvacua_difference_bis}
E_{\rm vac}^{\rm diff}(\pm 1)=+\frac{1}{N}\sum_{k\in
\mathbb{Z}_{N}}f_\pm\left(\frac{2\pi
k}{N}+\frac{\pi}{N}\right)-f_\pm\left(\frac{2\pi k}{N}\right),
\end{equation}
where
\begin{equation}
f_\pm(x)=\sqrt{\left(\pm
1-\cos x \right)^2+\gamma^2\sin^2 x }.
\end{equation}
In the thermodynamic limit, by applying the same technique used in
\eref{eq:diff_firt_derivative_in_thermod_limit}, equation
\eref{eq:XYvacua_difference_bis} becomes
\begin{eqnarray}
E_{\rm vac}^{\rm diff}(\pm 1) &\sim&
-\frac{1}{2\pi}\int_{0}^{2\pi}\left[f\left(x+\frac{\pi}{N}\right)-f(x)
\right]  \mathrm{d}x
\nonumber\\&\sim&
-\frac{1}{2\pi} \frac{\pi^2}{2!N^2}\int_{0}^{2\pi}f''(x){\rm
d}x
\sim \frac{\pi}{2 N^2}\gamma, 
\qquad N\to\infty,
\end{eqnarray}
where we used the equality $f'_{\pm}(0)=-\gamma$. See figure
\ref{fig:vacua_diff_g=pm1}.
\begin{figure}
\begin{center}
\includegraphics[width=0.7\columnwidth]{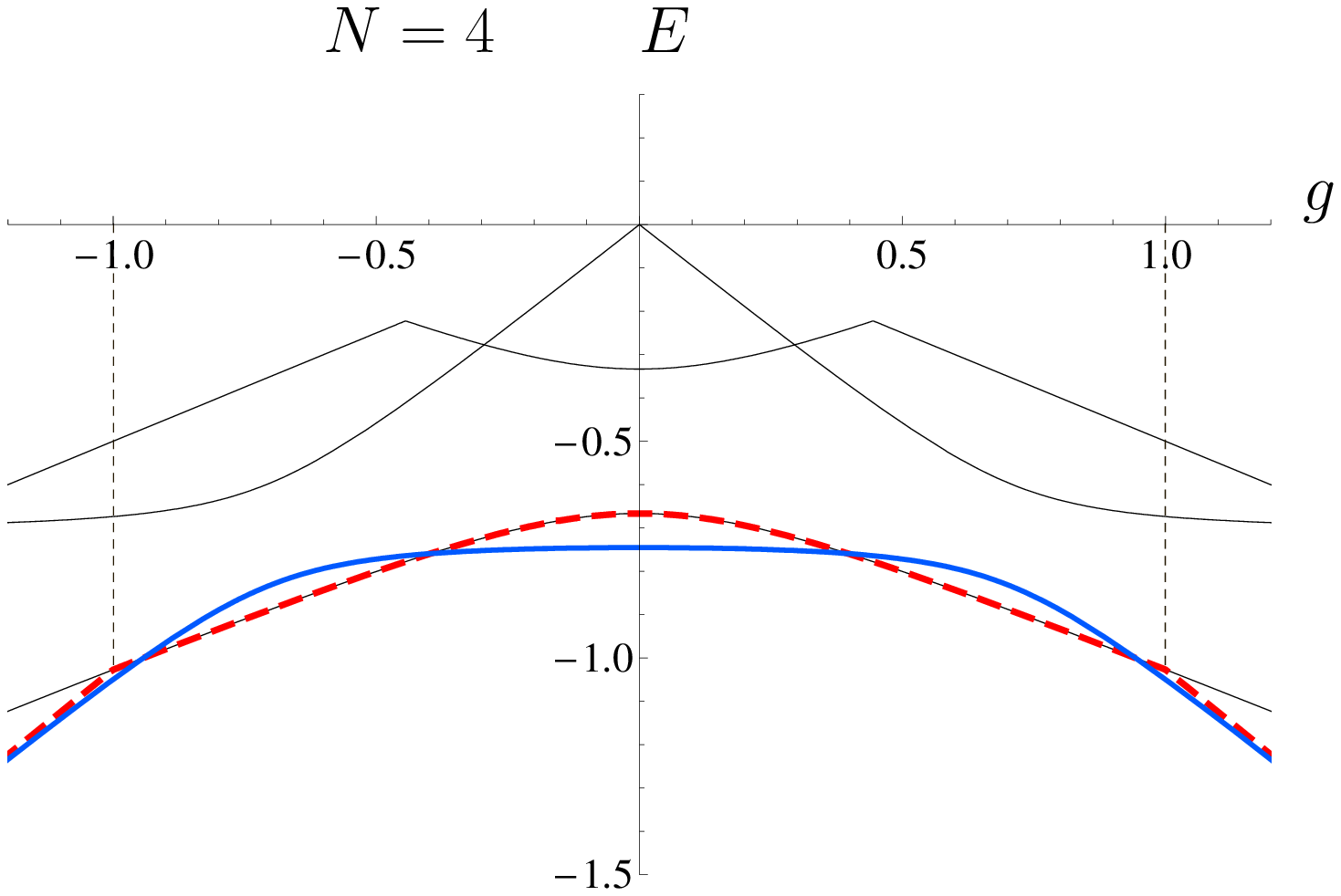}
\qquad\\
\includegraphics[width=0.7\columnwidth]{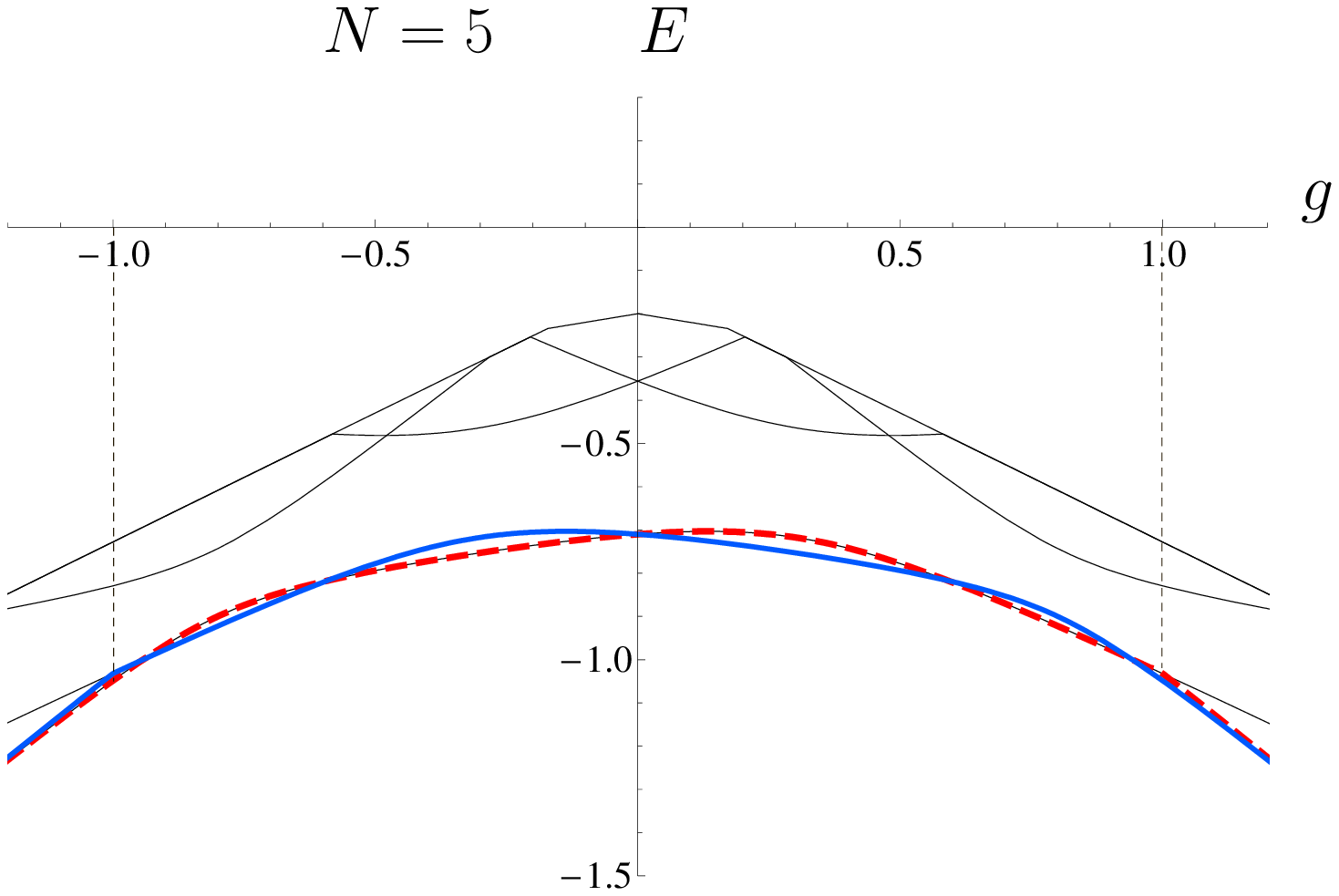}
\end{center}
\caption{The thin lines represent the low energy part of the spectrum of the XY
chain, and the thick solid and dashed lines refer to the two vacua
energy densities, $E_{\rm vac}^{(-)}$ and $E_{\rm vac}^{(+)}$, respectively:
these vacua alternatively coincide with the ground state and the
first excited state for $|g|\leq 1$. When $g=\pm1$ one vacuum energy
is the ground state energy, while the other one does not corresponds
to any physical level. The transition points are the forerunners of the quantum phase transition.} \label{fig:energySpectra}
\end{figure}
In \fref{fig:energySpectra} we display the low energy part of the spectrum (thin
lines) and the energy density  of the two vacua (thick lines): at
$g=\pm1$ the ground state is the winning vacuum that has no
singularities, the first excited level coincide with the losing
vacuum for $g\in(-1,1)$. Its second derivative diverges at $g=\pm
1$, forerunning the quantum phase transitions. Observe that they are
at the transition between a physical state, which coincides with the
first excited level, and an unphysical state, which does not
corresponds to any physical level: for $|g|>1$ the losing vacuum is
unphysical. Summarizing, we identify as forerunners of the quantum
phase transition those points of the losing vacuum energy density whose second derivative diverges.
These points are associated to single Bogoliubov fermions and belong to the crossing between the first excited level
and the unphysical vacuum for  finite size systems. When
$N\rightarrow \infty$ they approach the ground state as $N^{-2}$.

\subsection{Quantum phase
transitions in the XX model} \label{sec:XX Model QPTs} 
\begin{figure}
\begin{center}
\includegraphics[width=0.69\columnwidth]{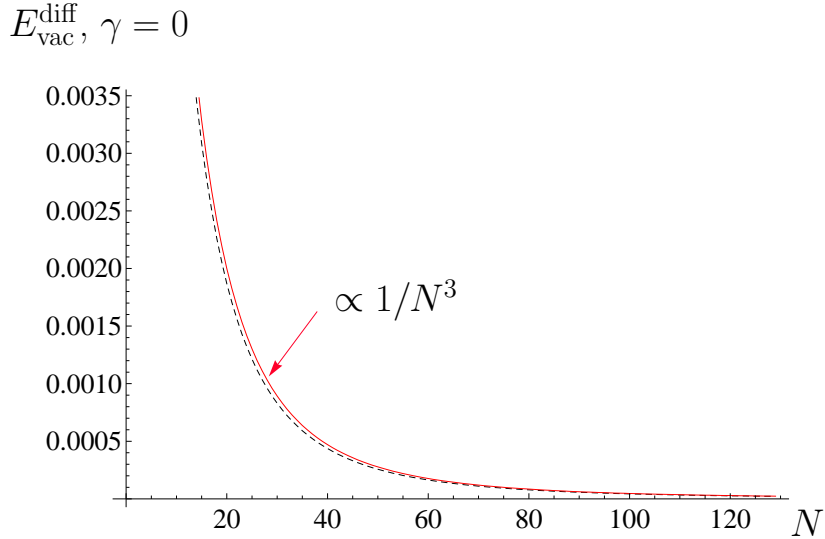}\caption{Difference between vacuum energies at $\ell=3$ versus $N$ (dashed line), and
its asymptotic approximation (solid line).
}\label{fig:xxvacuadifference}
\end{center}
\end{figure}

As observed
in Section \ref{sec:theXXmodel} the XX model ($\gamma=0$) is
characterized by the only presence of  single fermions, and the
absence of Bogoliubov pairs. As a result, all points
$g=\cos\left(\frac{2\pi (\alpha+k)}{N}\right)$ (in both parity
sectors) with $k \in \mathbb{Z}_{N}$  can be considered quantum
phase transitions forerunners.
\begin{figure}
\begin{center}
\includegraphics[width=0.7\columnwidth]{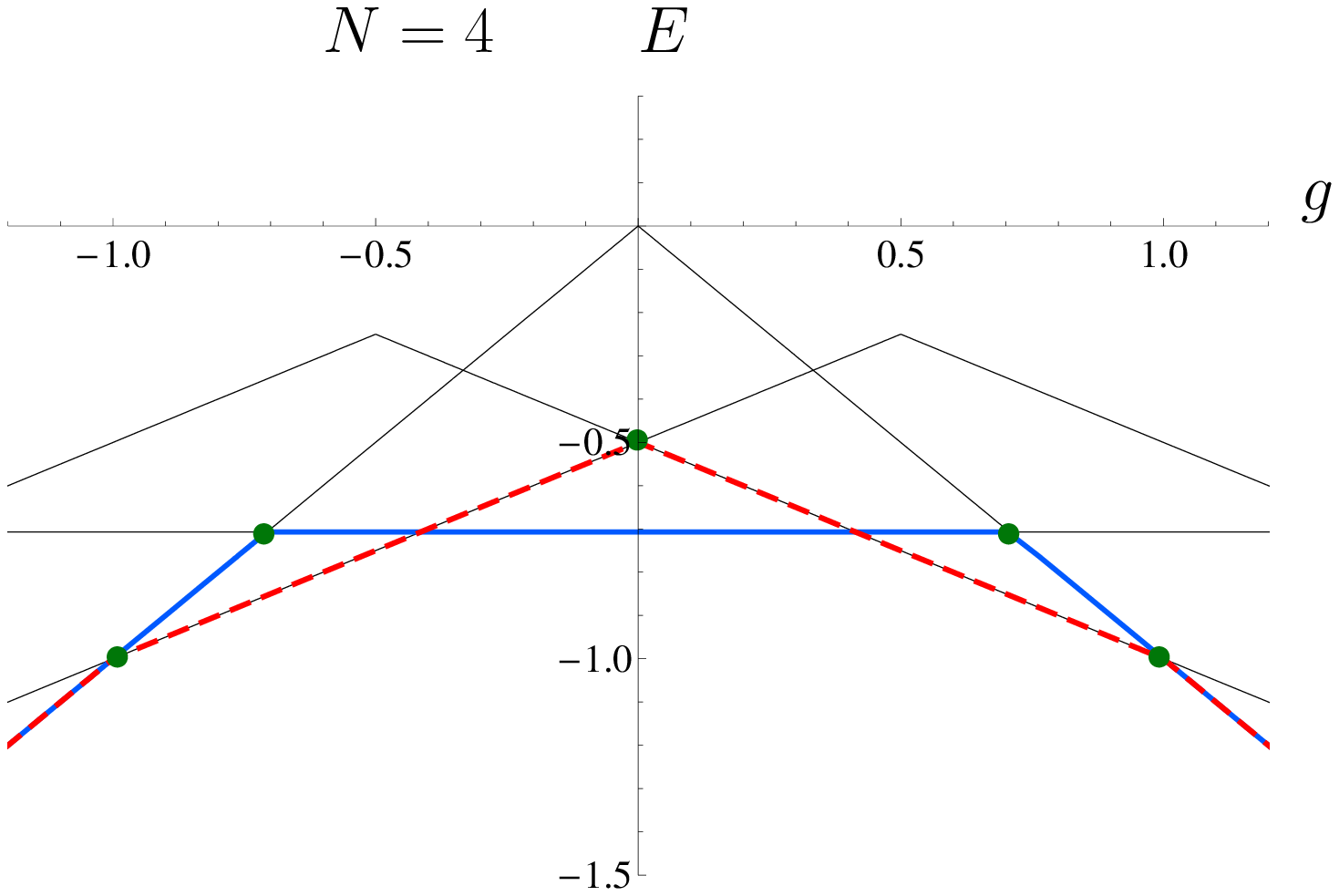}
\quad \quad\\
\includegraphics[width=0.7\columnwidth]{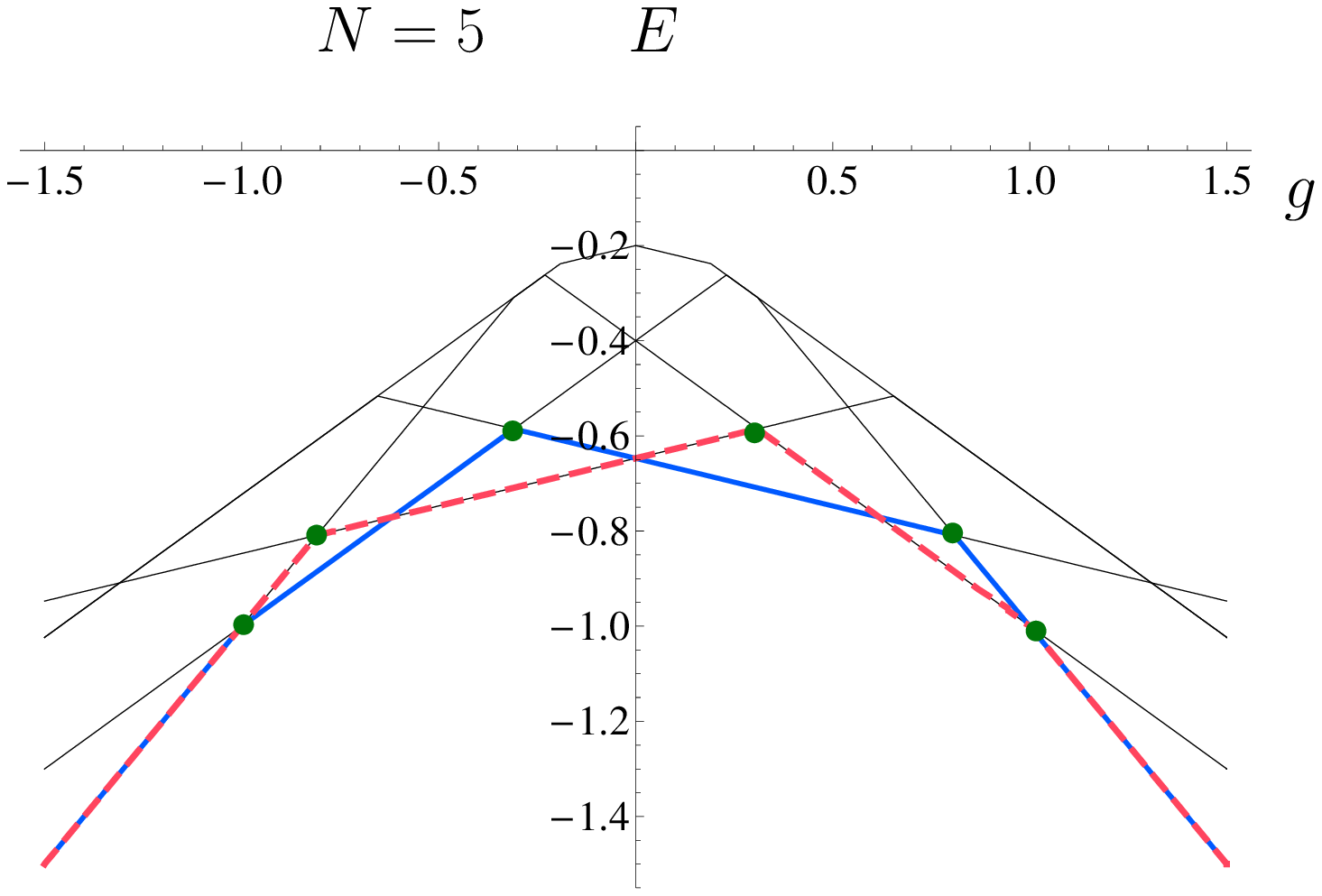}
\caption{The thin lines represent the low energy part of the spectrum of the XX
chain with $N=4$ and $N=5$ spins; the solid and dashed tick lines refer
the two vacua energy densities in the parity sectors with
$\varrho=-1$ and $\varrho=+1$, respectively.
The forerunners of the (continuous) quantum phase transition points are  indicated with
bold points;
they are given by $g_k=\cos\left(\frac{2 \pi k}{N}\right)$,
$k \in \mathbb{Z}_{N}$.}\label{fig:XX_spectrum_andQPTforerunners}
\end{center}
\end{figure}
See (\ref{eq:XXvacuum_alpha0})-(\ref{eq:XXvacuum_alpha1/2}) and
compare with
\eref{eq:vacuum_alpha0Neven}-\eref{eq:vacuum_alpha1/2Nodd}. Indeed,
the second derivative of the vacua energy density contains a Dirac
delta function at these points and, apart from $g=\pm1$, they all
belong to the first excited level like in  the XY model, see Fig.
\ref{fig:XX_spectrum_andQPTforerunners} (we will focus on $g=\pm1$
at the end of this section). In the thermodynamic limit these
points forerunning the quantum phase transition approach the ground
state, becoming critical points. Consider for example
$g_\ell=\cos\left(\frac{2\pi \ell}{N}\right)\neq \pm 1$; the energy difference
between the vacua is now given by
\begin{eqnarray}
\fl E_{\rm vac}^{\rm diff}(g_\ell)\!&=&\!-\frac{1}{N}\!\sum_{k\in\mathbb{Z}_{N} \backslash\{\ell\}}\!\!
\left[\left|\cos\left( \frac{2\pi
\ell}{N}\right)-\cos\left(\frac{2\pi
k}{N}\right)\right|-\left|\cos\left(\frac{2\pi
\ell}{N}\right)-\cos\left(\frac{2\pi
k}{N}+\frac{\pi}{N}\right)\right|\right]\nonumber\\
\fl &  & +\frac{1}{N}\left|\cos\left(\frac{2\pi
\ell}{N}\right)-\cos\left(\frac{2\pi
\ell}{N}+\frac{\pi}{N}\right)\right|.
\end{eqnarray}
By using the same technique of the previous section one gets
\begin{eqnarray}
\fl \qquad E_{\rm vac}^{\rm diff}(g_\ell) &=&
\frac{1}{2\pi}\int_{0}^{2
\pi}\left[\frac{\pi}{N}f'_\ell(x)+\left(\frac{\pi}{N}\right)^2
\frac{f''_\ell(x)}{2!}+\left(\frac{\pi}{N}\right)^3
\frac{f'''_\ell(x)}{3!}+O\left(\frac{1}{N^4}\right)\right]\mathrm{d}x
\nonumber\\
& &+\frac{1}{N}
\left|\cos\left(\frac{2\pi\ell}{N}\right)-\cos\left(\frac{2\pi\ell}{N}+\frac{\pi}{N}\right)\right|,
\qquad N\to\infty
\label{eq:XXvacua_diff_in_thermodynamic_limit}
\end{eqnarray}
where
\begin{equation}
f_\ell(x)=\left|\cos\left(\frac{2\pi
\ell}{N}\right)-\cos x
\right|.
\end{equation}
From the symmetries of $f(x)$ and its derivatives, it follows that
equation \eref{eq:XXvacua_diff_in_thermodynamic_limit} becomes
\begin{equation}
\fl \quad E_{\rm vac}^{\rm diff}(g_\ell) \sim \frac{1}{N}\sqrt{\left(\cos\left(\frac{2\pi\ell}{N}\right) -\cos\left(\frac{2\pi\ell}{N}+\frac{\pi}{N}\right)\right)^2}
\sim 
\frac{ 2 J \pi^2 \ell}{N^3}, \qquad N\to \infty.
\end{equation}
See \fref{fig:xxvacuadifference}.
Therefore, in the thermodynamic limit the forerunners of the
quantum phase transition in the isotropic XX model  approach the
ground state faster than the ones of the XY model (with $\gamma \neq
0$). Compare figures \ref{fig:vacua_diff_g=pm1} and
\ref{fig:xxvacuadifference}.

As shown in \fref{fig:XX_spectrum_andQPTforerunners},  the
intersection points of the two vacua (which coincide with the level
crossing points $g_{l}(n)$ discussed in \sref{sec:theXXmodel}) are
characterized by a discontinuity of the first derivative for finite
size chains.
By deriving the energy difference (\ref{eq:diff}), one
can show that the
discontinuity of the first derivative at the points of level crossing
scales like
$1/N$; 
therefore in the thermodynamic limit the divergence of the second derivative vanishes, as for the XY Hamiltonian with $\gamma \neq 0$.

Let us finally consider the points $g=\pm1$: on one hand they are
level crossing points ($g=\pm\sqrt{1-\gamma^2}$, $\gamma=0$), on the
other hand, following the same criterion introduced for the XY
model, they can be considered as forerunners of quantum phase
transitions: what happens in this particular case is that these
points belong to the ground state already for finite $N$. Another
crucial difference between the anisotropic case and the XX model is
that, since all Bogoliubov fermions are single,  there are $N+1$ points forerunning the quantum phase transition.
 Thus in the $N\to \infty$ limit they densely fill the
interval  $[-1,1]$ of $g$  
and yield, as one expects \cite{ModelliEsatti1}, a
continuous quantum phase transition in this interval.

\section*{Conclusions}
In this paper we analyzed the XY model with periodic boundary
conditions. Being interested in finite size systems, we did not
neglect the boundary term which derives from the Jordan-Wigner
transformation. In order to diagonalize the Hamiltonian we deformed
the discrete Fourier transform with a local gauge coefficient
depending on the parity of spins down, anti-parallel to the magnetic
field. We then showed that in the Fourier space there are two
classes of fermions, single and coupled ones; this distinction is
crucial in order to determine the Bogoliubov transformation, which
is also gauge dependent. From the expression of the diagonalized
Hamiltonian we reinterpreted the ground state and the first excited
level of the system as given by a competition between the vacuum
energies of the two parity sectors. We finally introduced a
criterion to find those values of the magnetic field that can be
considered forerunning quantum phase transitions in the
thermodynamic limit. They are associated to single Bogoliubov fermions and to the level crossings
between physical and unphysical states.

There is considerable
interest in the study of entanglement for quantum spin chains, both
in view of applications and because of their fundamental interest. See, for example, the results concerning the XX chain \cite{plastina,CFFP,XX Model}.
Future activity will focus on the study of the properties of the
multipartite entanglement of the ground state in terms of the
distribution of bipartite entanglement \cite{scott,MMES} and on the investigation of the possible connections with quantum phase transitions in
the thermodynamic limit.

\ack
We thank Saverio Pascazio for a critical reading of the manuscript.
This work is supported by the European Community through the Integrated Project EuroSQIP.

\end{document}